# *Competitive Balance measures and the Uncertainty of Outcome Hypothesis in European Football*

For Mathsport 2019 special issue

V. MANASIS[1],
*School of Physical Education and Sport Science, National and Kapodistrian University of Athens, Greece*
I. NTZOUFRAS
*Department of Statistics, Athens University of Economics and Business, Greece*
J.J. READE
*Department of Economics, University of Reading, UK*

Competitive balance is an important concept for professional sports and one of the key issues that European football has to address in order to ensure its long-term prosperity. However, the quantification of competitive balance is not an easy task. The difficulties are mainly associated with its multi-dimensionality character as well as the structure of each particular sport. This article uses data from eight domestic leagues over 60 years to identify the best index for a holistic view of competitive balance in European football. The findings support the longstanding *Uncertainty of Outcome Hypothesis* using indices designed for the important three identified levels of competition and offering a weighting pattern for ranking places. Important conclusions may be derived concerning the relative importance of different aspects of competitive balance depending on the specific features of the best index.

**Keywords**: competitive balance; uncertainty of outcome hypothesis; European football; professional team sports; attendance; demand equation

---

[1] Corresponding author. Email: vmanasis@phed.uoa.gr

# *Competitive balance measures and the Uncertainty of Outcome Hypothesis in European football*

## *Abstract*

*Competitive balance is an important concept for professional sports and one of the key issues that European football has to address in order to ensure its long-term prosperity. However, the quantification of competitive balance is not an easy task. The difficulties are mainly associated with its multi-dimensionality character as well as the structure of each particular sport. This article uses data from eight domestic leagues over 60 years to identify the best index for a holistic view of competitive balance in European football. The findings support the longstanding Uncertainty of Outcome Hypothesis using indices designed for the important three identified levels of competition and offering a weighting pattern for ranking places. Important conclusions may be derived concerning the relative importance of different aspects of competitive balance depending on the specific features of the best index.*

### 1. Introduction

The main aim of this paper is to identify which competitive balance measure is most closely associated with the attendance of spectators. Our approach can offer a holistic view of competitive balance in eight major and minor leagues for a timeline of 60 years. By this way, we are trying not only to empirically test the "Uncertainty of Outcome Hypothesis" (*UOH*, Fort & Maxcy, 2003) but also to identify which measure expresses more efficiently the essence of this theory. To be more specific, *UOH* states that the more competitive is a sports league, the more fans' interest it stimulates. The fans' stimulation in this study is measured from the number of tickets (attendance) which is considered as the standard measure for the fans behavior according to the reviews on sport demand by Borland & MacDonald (2003) and Villar & Guerrero (2009). By identifying statistically significant competitive balance measures which are negatively associated with attendance, we can assume that the *UOH* is confirmed (at least in the level of attendance). By identifying which measures are good predictors of the attendance, we can conclude which measures are the most appropriate to explain and predict the interest of the fans.

Before moving forward, we need to state that the attendance only partially captures the overall fans' interest nowadays. Admittedly, there are also other sport features such tv or streaming viewership and team product sales that also reflect the interest of fans for a professional team or league more generally. But this response was used mainly for two reasons. First, its importance is unquestionable through time. Secondly, attendance is readily available from various sources with small measurement errors even for early seasons of the game. Moreover, other measurements of interest (such as tv viewership, streaming measures or sales) have been developed later in time and therefore the incorporation of such measures in an empirical study will not be feasible for older seasons.

Many different approaches have been introduced to measure competitive balance; as Zimbalist (2002, p. 112) puts it: "*there are almost as many ways to measure competitive balance as there are to quantify money supply*". A possible explanation is the fact that the concept of competitive balance is multi-faceted and various factorizations have been emerged in the literature (see Sloane, 1971; Jennett, 1984; Cairns, 1987; Vrooman, 1996; Fort, 2003; Szymanski, 2003; Pawlowski & Nalbantis, 2019). The present study adopts the popular three-dimensional factorization of competitive balance (e.g. Quirk & Fort, 1997; Czarnitzki & Stadtmann, 2002; Borland & MacDonald, 2003; Brandes & Franck, 2007). The three dimensions under consideration are the following: a) the *match uncertainty*, which refers to a particular game, b) the *seasonal* dimension, which focuses on the relative quality of teams in the course of a particular season and c) the *between-seasons* dimension, which focuses on the relative quality of teams across seasons.

There is no universally accepted measure or index of competitive balance. Different measures of competitive balance may be assumed optimal (in the sense that best describe the structure and the uncertainty) for different sports, leagues or competitions. (Zimbalist, 2003). Our focus is on European football (soccer), which, according to Gerrard (2004, p. 39) is the heartland of football, the only truly global team sport. However, European football leagues are complex in structure, in that domestic championships are multi-levelled tournaments offering multiple prizes as opposed to the common single prize offered by North American ones (Kringstad & Gerrard, 2007). On this study, attention is given to the three-levelled structure of European football introduced by Manasis, Avgerinou, Ntzoufras, and Reade (2013). This sub-competition structure can be also related to the playoff uncertainty (Fort & Lee, 2006), the Champions League qualification uncertainty (Pawlowski & Anders, 2012), and the competitive intensity to different stakes (Scelles *et al.*, 2013a, Scelles, 2017).

The attention of the economic analysis of competitive balance is the effect on the fans' behaviour, which is the longstanding "*Uncertainty of Outcome Hypothesis"* (*UOH*, Fort & Maxcy, 2003). Based on the empirical studies, the relationship between competitive balance and fans' interest, exhibited by their demand for league products, is a matter of debate. The majority of studies testing the *UOH* have focused on the relevance of match uncertainty (Pawlowski & Nalbantis, 2019). However, the evidence of a positive effect of match uncertainty on demand is either relatively weak or even contradictory (see Borland & MacDonald, 2003; Pawlowski, 2013; Coates, Humphreys, & Zhou, 2014). This has motivated further reasearch leading to the concepts of reference-dependent preferences and loss aversion (Coates, Humphreys, & Zhou, 2014), the concept of competitive intensity (Scelles *et al.*, 2013a, 2013b; Scelles, 2017) as well as subjective evaluations of competitive balance (Pawlowski, 2013; Pawlowski & Budzinski, 2013, 2014). According to Borland and MacDonald (2003), the evidence of an effect of the seasonal and between-seasons dimensions on demand is more clear (e.g. Hunt & Lewis, 1976; Schmidt & Berri, 2001; Humphreys, 2002; Lee, 2004).

Therefore, the present study focuses on these two longer time-dependent dimensions, which are of utmost importance for the study of competitive balance. In the first level, we are interested to test for the *UOH* by investigating the relation of completive balance measures with the attendance while, in the second level, and assuming that *UOH* is true, our objective is to identify the competitive balance index which is most associated with this response. Following Zimbalist's (2003) assumption, we argue that the impact on the number of spectators should be used as a filter among potential competitive balance indices. Additionally, key conclusions may be derived concerning the relative importance of different aspects of competitive balance depending on the specific features of the best index.

The effect of competitive balance on attendance in European football is estimated via an empirical study with data from several domestic leagues for an extended period of seasons. Our dataset enables us to implement advanced statistical methods that lead us to solid conclusions arising from the statistical analysis. After presenting several indices measuring competitive balance, the article proceeds with variables description and the methodology followed for the construction of the statistical model. The article proceeds with the empirical results, the discussion of the main findings and finally some concluding remarks derived from the analysis.

## 2. Indices of Competitive Balance

The increase of existing competitive balance indices indicates progress towards a more efficient and accurate quantification of this notion in sports. However, the existence of a wide variety of different interpretations of competitive balance (Michie & Oughton, 2004) generates a difficulty to identify a measure that fully captures all its characteristics. For this reason, as Zimbalist (2002) notes, a great diversity of different approaches has been introduced. In the current study, we have selected normalized indices with clear bounds corresponding to the extreme cases of perfect balance to complete imbalance in order to account for differences in the size of the league $N$ across seasons and leagues. Finally, the selected between-seasons indices should be able to handle the promotion-relegation rule which creates complications due to the changes of the participating teams[1].

Seasonal indices measure the relative quality or strength of teams during a particular season. One most widely used is the so-called *National Measure of Seasonal Imbalance* (*NAMSI*), introduced by Goossens (2006), which compares the observed standard deviation of winning percentages with the standard deviation in the case of a completely unbalanced league (i.e. the most undesirable one). Since competitive balance is essentially concerned with inequalities among teams some specific indices have been adopted from the industrial organisation theory. The *Herfindahl-Hirschman Index* (*HHI**), introduced by Depken (1999) and standardized by Owen *et al.* (2007), is often used to illustrate the distribution of a variable by measuring its degree of concentration across units. Also, the *Adjusted Gini Coefficient* (*AGini*), introduced to the sports context by Schmidt (2001) and Schmidt and Berri (2001), and developed by Utt and Fort (2002), is a traditional index of inequity measuring the degree to which a championship deviates from equality.

[1] Manasis & Ntzoufras (2014) offer a detailed discussion concerning the applicable indices to European football.

A set of specially designed indices has been introduced by Manasis *et al.* (2013) to account for the three levels of competition in European football leagues in which teams compete for the corresponding ordered sets of prizes or punishments.

a) The first level stands for the championship title.
b) The second level stands for the $K$ qualifying places for participation in European tournaments the following season.
c) The third level stands for the $I$ relegation places.

Based on the assumption that the overall degree of competitive balance is determined by the corresponding degrees of competitiveness in the three league levels, Manasis *et al*. (2013) employed a simple ensemble or index averaging technique to quantify the overall competitive balance. More specifically, they developed the *Normalised Concentration Ratio for the Champion* ($NCR_1$), the *Adjusted Concentration Ratio* ($ACR_K$), and the *Normalised Concentration Ratio for Relegated Teams* ($NCR^I$) which capture competitiveness at the first level, at the first two levels (combined), and the third level, respectively. As a by-product, they also defined, the *Special Concentration Ratio* ($SCR_K^I$), which is an ensemble measure of all the previous ones and captures overall competitiveness for all three levels. The comprehensive $SCR_K^I$ offers a benchmark weighting scheme in which the top $K$ ranking places receive the highest weight (which is decreasing when moving downwards to the league table) while the bottom $I$ league positions, that usually lead to relegation, receive weights lower than the top $K$ teams but higher than the middle ranking places.

Between-seasons indices refer to the longest time-wise dimension and measure the relative quality of teams across consecutive seasons (at least two) by considering the ranking fluctuations as the main unit of measurement. Hann, Koning, and van Witteloostuijn (2007) propose the *index of Dynamics* ($DN_t$) to measure the ranking mobility of all teams across two consecutive seasons by summing the absolute ranking differences. Also, Groot (2008) introduces in sports the statistical index *Kendall's tau coefficient* ($\tau$) which illustrates the overall ranking turnover within a league between two seasons. Lastly, Manasis and Ntzoufras (2014), following the methodology of Manasis *et al*. (2013) for the seasonal indices, have created a set of corresponding between-seasons three-levelled indices; the *Dynamic Index for the Champion* ($DN_1$), the *Adjusted Dvnamic Index* ($ADN_K$), the *Dynamic Index for Relegated Teams* ($DN^I$), and the *Special Dynamic Index* ( $SDN_K^I$ ). Interestingly enough, seasonal and between-seasons three-levelled indices have the same range and scale. Moreover, the weighting scheme of the league positions is identical for a given league. A detailed description of the special designed indices both for the seasonal and the between-seasons dimensions is presented at the Appendix.

By virtue of these properties, a group of bi-dimensional *Dynamic Concentration* indices is introduced here to capture both the seasonal and the between-seasons dimension. Essentially, a *Dynamic Concentration* index employs the specific qualities of a *Normalised Concentration Ratio* (seasonal dimension) as well as a *Dynamic Index* (between-seasons dimension) by simply averaging the corresponding indices. The new set of ensample bi-dimensional indices are the following:

1. the *Dynamic Concentration index for the Champion* $DC_1=(NCR_1+DN_1)/2$
2. the *Adjusted Dynamic Concentration* $ADC_K = (ACR_K+ADN_K)/2$
3. *the Dynamic Concentration for Relegated Teams* $DC^I= (NCR^I+DN^I)/2$
4. *the Special Dynamic Concentration 3-levelled Index* $SDC_K^I =(SCR_K^I +SDN_K^I )/2$

In the following sub-section we present some details of these bi-dimensional indices while some summary details about the three-levelled seasonal and between-seasons competitive balance can be found at the Appendix and at the original publications of Manasis *et al.* (2013) and Manasis & Ntzoufras (2014).

### **2.1** *Dynamic Concentration Index for the Champion*

The *Dynamic Concentration index for the Champion* ($DC_1$) nicely depicts the degree of the champion's domination both seasonally and dynamically. The calculation of $DC_1$ is derived by the average of its corresponding component (single dimensional) indices $NCR_1$ and $DN_1$:

$$DC_1 = \frac{NCR_1 + DN_1}{2} = \frac{P_1 - 2r_1}{4(N-1)} \quad \textbf{(1)}$$

where $P_1$ stands for the number of points, $r_1$ stands for the absolute ranking difference across two consecutive seasons, and $N$ stands for the number of teams in the league. In essence, $DC_1$ is a bi-dimensional index, which portrays the champion's overall behaviour in terms of competitive balance. The index is interpreted as:

a) The domination degree by the champion (seasonal dimension).
b) The degree of ranking mobility or dynamic domination by the champion (between-seasons dimension).

### **2.2** *Adjusted Dynamic Concentration for the Top-K Teams*

The *Adjusted Dynamic Concentration* ($ADC_K$) summarises the behaviour of the top $K$ teams that qualify in any European tournament. Similarly, to the $DC_1$, the calculation of the index is merely the average of the corresponding $ACR_K$ and $ADN_K$ indices:

$$ADC_K = \frac{ACR_K + ADN_K}{2} = \frac{1}{2K}\left[ w_i \sum_{i=1}^{K} (P_i - 2r_i) - C_K \right] + \frac{1}{2} \quad \textbf{(2)}$$

where $K$ stands for the qualifying places for participation in European tournaments the following season, $w_i$ stands for the weight attached to the $i$th team and $C_K$ is a constant term. The weights $w_i$ and the constant $C_K$ are identical to the corresponding ones used to define the seasonal $ACR_K$ index; for details, see equations 10 to 13 at the Appendix.

Since $ADC_K$ refers to several single-dimensional component indices, its interpretation is not simple and is given as follows:

a) The degree of concentration or domination by the top $K$ teams and the degree of competition among the top $K$ teams (seasonal dimension).
b) The degree of ranking mobility or dynamic domination by the top $K$ teams and the degree of ranking mobility or dynamic competition among the top $K$ teams (between-seasons dimension).

### **2.3** *Dynamic Concentration for Relegated Teams*

The *Dynamic Concentration for Relegated Teams* ($DC^I$) captures the bi-dimensional performance of the $I$ relegated teams. $DC^I$ effectively depicts the behaviour of the last $I$ teams, since it borrows its specific features from its component $NCR^I$ and $DN^I$ indices. Following the previous discussion, the formula of the index is given by:

$$DC^I = \frac{NCR^I + DN^I}{2} = \frac{2N - I - 1}{2(N-I)} - \frac{\sum_{i=N-I+1}^{N} (P_i + 2r_i)}{4I(N-I)} \quad \textbf{(3)}$$

where $I$ stands for the number of relegation places. The bi-dimensional $DC^I$ index is interpreted as:

a) The degree of competition for relegation or the degree of weakness of the $I$ relegated teams (seasonal dimension).
b) The degree of dynamic competition for relegation or dynamic weakness of the $I$ relegated teams (between-seasons dimension).

### *2.4 Special Dynamic Concentration 3-levelled Index*

The *Special Dynamic Concentration* ($SDC_K^I$) is an ensemble index as it captures all three levels in both the seasonal and the between-seasons dimensions. More specifically, $SDC_K^I$ reveals the bi-dimensional behaviour both of the top $K$ and the bottom $I$ teams. It is calculated by simply averaging the corresponding $SCR_K^I$ and $SDN_K^I$ indices as follows:

$$SDC_K^I = \frac{SCR_K^I + SDN_K^I}{2} \Leftrightarrow$$

$$SDC_K^I = \frac{1}{2(K+1)}\left( w_i \sum_{i=1}^{K} (P_i - 2r_i) - w_I \sum_{i=N-I+1}^{N} (P_i + 2r_i) - C_K + C_I \right) + \frac{1}{2} \quad \textbf{(4)}$$

for $I \leq N/2$, $K \leq N/2$, $I+K<N$

where $w_I$ stands for the weight attached to the bottom $I$ teams and given by $1/(2I(N–I))$, $C_I$ is a constant term given by $(N–1)/(N–I)$.

As it is expected, the interpretation $SDC_K^I$ is not straightforward, since it possess three different qualities in two dimensions:

a) The degree of concentration or domination by top $K$ teams, the degree of competition among the top $K$ teams and the degree of competition for relegation or the degree of weakness of the $I$ relegated teams (seasonal dimension).
b) The degree of ranking mobility or dynamic domination by the top $K$ teams, the degree of ranking mobility or dynamic competition among the top $K$ teams and the degree of dynamic weakness of the $I$ relegated teams or the degree of dynamic competition for relegation.

The weighting pattern offered by the index, for realistic values of $K$ and $I$, rates the top $K$ teams with a decreasing pattern of weights which are higher than the corresponding weights of the bottom $I$ teams. Similarly to the corresponding uni-dimensional three levelled indices, any of the $I$ positions of the leagues is rated higher than the teams in the middle of the ladder ($N–K–I$) since those are not included in the index. This weighing pattern is not necessarily an optimal one, but it provides an intuitively simple and plausible benchmark for the study of competitive balance in European football.

## 3. Data and Variables

We work within the *UOH* framework in order to determine the relative significance of each index described in the previous section. To our knowledge, the investigation of aggregate attendance across different leagues or countries has received limited attention in the relevant literature, see for example the studies of Lee (2004) and Schmidt and Berri (2001); however, none of these studies concerns competitive balance in European football.

Data selection is based: a) on the availability for a long period of 60 seasons, b) representation from southern, central and northern Europe, and c) representation from both top five and small countries in terms of their total football revenues. The starting season coincides with the establishment of the highest league in Greece. We have collected data over eight different European leagues (Belgium, England, France, Germany, Greece, Italy, Norway and Sweden) for the seasons 1959/60-2018/19 with the exception of Belgium, Germany and Norway starting from 1966/67, 1963/64, and 1962/63 respectively (Table 1)[2]. Therefore, the collected dataset consists of an unbalanced panel dataset with eight cross units ($n$) (European domestic leagues) over 53 to 60 seasons ($T$). This is the largest dataset in terms of both the number of seasons and countries that have appeared in the competitive balance literature according to our knowledge.

[2] 1963 is the founding year for the German football league (Bundesliga) and the Norwegian single top division (1. Divisjon). The first one-league season in Norway played in spring-autumn (season 1962/63 in our data).

For each year and country, the final league table results (points and ranking, which are transformed to competitive balance indices), the total number of tickets and a collection of covariates are collected[3]. The number of tickets is referred to as the annual average attendance per game (in order to also account for the size of the league) for each season and it is used in the log-scale (denoted in the following as ln*ATT*). Attendance at league games is a conventional measure for the fans' behaviour based on the most complete reviews for demand in professional team sports (Borland & MacDonald, 2003; Villar & Guerrero, 2003, Villar & Guerrero, 2009). As a referee pointed out, it would have been more appropriate to use the average percentage of capacity utilization. However, such data were not easily accessible for all 8 leagues and for all of the seasons under consideration.

TABLE 1. Dataset Basic Characteristics

| *Country* | *Starting season* | *Ending season* | *Total seasons* |
|---|---|---|---|
| Belgium (BEL) | 1966/67 | 2018/19 | 53 |
| England (ENG) | 1959/60 | 2018/19 | 60 |
| France (FRA) | 1959/60 | 2018/19 | 60 |
| Germany (GER) | 1963/64 | 2018/19 | 56 |
| Greece (GRE) | 1959/60 | 2018/19 | 60 |
| Italy (ITA) | 1959/60 | 2018/19 | 60 |
| Norway (NOR) | 1962/63 | 2018/19 | 57 |
| Sweden (SWE) | 1959/60 | 2018/19 | 60 |

The covariates we consider in the analysis are the following:

- ln*CB*: Index of competitive balance (log scaled)
- ln*POP*: National population (log scaled)
- ln*RGNI*: Real per capita gross national disposable income (log scaled)
- *Un*: Unemployment rate
- *d*97: Dummy variable for the period after season 1997/98
- *Trend*: Time trend

Given that a suitable form is important to the analysis (Villar & Guerrero, 2009), the employed natural logarithm for *CB*, *POP* and *RGNI* allows for non-linear (exponential) relationship of the explanatory with the response variable and offers the economically important elasticity interpretation. Testing the significance of a large number of indices is quite innovative compared with the common practice; Lee and Fort (2008) employ four indices (or factors) whereas Humphreys (2002) and Lee (2004) employ three. Here we have included in our analysis seven seasonal, six between-seasons and four bi-dimensional competitive balance indices. In particular, we have tested for the significance of the following indices:

a) ***Seasonal indices*** (7 in total): *NAMSI*, *HHI**, *AGini*, $NCR_1$, $ACR_K$, $NCR^I$, and $SCR_K^I$

b) ***Between seasons indices*** (6 in total): $DN_t$, $\tau$, $DN_1$, $ADN_K$, $DN^I$, and $SDN_K^I$, and

c) ***Bi-dimensional indices*** (4 in total): $DC_1$, $ADC_K$, $DC^I$, and $SDC_K^I$

The main explanatory variable is the index of competitive balance (ln*CB*) for which a negative sign in the coefficient will support the *UOH*, since the value of all indices ranges from zero (perfect balance) to one (complete imbalance)[4].

The number of top *K* teams on the above special designed indices varies across seasons and leagues according to the performance of the teams in the European leagues and the number of qualification places awarded by UEFA (the rules for the distribution of these qualification places across countries are also revised by UEFA from time to time).The number of relegated teams *I* usually remains stable for specific time periods with sporadic changes according to the strategic decisions of each national domestic football association about the formation of the league. In some cases where exceptionally low *K* and/or *I* (in comparison to the usual values) is observed, a different approach is followed in order to avoid over-penalizing specific league seasons. To avoid this over-penalization, we employed a lower limit for both *K* and *I*[5]. For a characteristic example we may refer to the 5-seasons period (1986-1990) that UEFA banned English football where we considered as *K=5* as a legitimate number of top teams that attract fans' interest.

[3] For the calculation of the seasonal indices, any deducted points (e.g. due to misconduct of club officials or supporters, match-fixing) are awarded in order to depict the true sporting performance of the club. The official ranking is taken into consideration the following season for the calculation of the dynamic indices.

[4] Note that we use the rescaled version of $DN_t$ and $\tau$ (see Manasis & Ntzoufras, 2014).

Based on the standard consumer theory model, price is an important economic factor; however, relevant data for such a large data panel is unavailable. Another important economic factor is the size of the potential market for which population is used as a proxy in a number of related demand studies (Jennett, 1984; Wilson & Sim, 1995; Rivers & DeSchriver, 2002; Donihue, Findlay, & Newberry, 2007). In our case, the employed national population (ln*POP*) is expected to be positively related with attendance.

Fans' buying power also constitutes an important economic factor, provided that attendance at football games is a normal good. The real deflated gross national disposable income per capita (ln*RGNI*) is a typical way to evaluate the income variable; Bird (1982) uses real consumption spending, Schollaert and Smith (1987) use household income, and Simmons (1996) uses regional real earnings. All else being equal, ln*RGNI* will positively affect attendance.

The macroeconomic factor of unemployment rate (*Un*) is also included as Borland and MacDonald (2003) suggest that attendance at sporting events may constitute a social outlet for unemployed persons. In periods of high unemployment, football games may become more popular to help people manage personal disappointment (Dobson & Goddard, 1996; Borland & Lye, 1992). However, the most common results in the literature (Villar & Guerrero, 2009) suggest that higher unemployment decreases attendance to stadiums.

The dummy variable for the period after season 1997/98 (*d*97) is included to account for the important structural changes of 'Bosman' case and Champion's League reform. The choice of season 1997/98 allows for these changes to have an effect on European football. Lastly, for a more reliable interpretation of the results a trend (*Trend*) is also included to capture other factors that affect demand for attendance at football games that change systematically over the seasons.

## 4. Methodology and Statistical Model

The nature of the dataset (small *n* and large *T*) stresses the adoption of panel unit root tests as described by Granger and Newbold (1974). The appropriate unit root test is the cross-sectionally augmented IPS (CIPS), proposed Pesaran (2007), which is a second-generation test appropriate for contemporaneously correlated errors. As was expected, all indices are stationary since competitive balance must be a self-correcting mechanism. According to Table 2, the observed values of CIPS are higher than the corresponding critical values at the 10% significance level (i.e. all p-values > 0.10). Therefore, the null hypothesis of a unit root process is not rejected for the examined economic variables of Table 2. Given that the dependent variable (ln*ATT*) is non-stationary, there is danger of spurious economic relationships (Phillips, 1986; Sims, Stock, & Watson, 1990). Differencing is one of the methods to solve spuriousness, however; the interpretation of the results becomes quite problematic since it cancels out any meaning of elasticity and information on levels. For that reason, we follow, an alternative solution, the autoregressive distributed lag relation (*ADL*) which allows a reliable estimation of the standard errors (Banerjee, Dolado, Galbraith, & Hendry, 1993; Hendry & Nielsen, 2007; Hendry & Doornik, 2009).

TABLE 2: Observed Statistic Values for Cross-sectionally augmented IPS (CIPS) Unit Root Tests

| *Variable* | *Constant* | | | *Constant & Trend* | | |
|---|---|---|---|---|---|---|
| ln*ATT* | -1.699 | | | -2.451 | | |
| ln*POP* | 0.563 | | | 0.025 | | |
| ln*RGNI* | -1.707 | | | -2.012 | | |
| *Un* | -1.559 | | | -1.548 | | |
| | *α=10%* | *α=5%* | *α=1%* | *α=10%* | *α=5%* | *α=1%* |
| Critical Values | < -2.05 | < -2.16 | < -2.37 | < -2.68 | -2.82 | -3.10 |

[5] We have used the lower thresholds of three and four for leagues of size $N \leq 16$ and $N \geq 18$, respectively, while for *I* we have used the lower threshold of two.

The most commonly used techniques for the analysis of panel data are the fixed and the random effect models (Hsiao, 2003; Baltagi, 2005). However, our long and narrow type of panel data requires a different approach; Greene (2008) offers several solutions on this topic. Since the interest is the interpretation of the results at European level, we follow the suggestion proposed by Kennedy (2008) by pooling the eight equations (one for each country) in order to improve efficiency. Estimating several equations together improves efficiency only if there are some restrictions on parameters (Hill, Griffiths, & Lim, 2008). The equation of our model takes the form:

$$A(L)\ln ATT_{it}=C_i+B_1(L)\ln CB_{it}+B_2(L)\ln POP_{it}+B_3(L)\ln RGNI_{it}+B_4(L)\ln UN_{it} \\ +B_5 d97_t+\sum_{g=1}^{m} B_{(5+g)i}\, Trend^{g}+\varepsilon_{it} \quad \textbf{(5)}$$

where $C_i$ is the constant of the model, $i$ stands for the country, $t$ stands for the year, $m$ stands for the degree of the trend variable, $L$ denotes the lag operator, and $\varepsilon$ is the error of the model, which is presumed to be white noise. The variation of the constant term $C_i$ in (5) allows for countries' heterogeneity and stands for the $i$th country-specific effect, which is mostly influenced by market and football factors related to football popularity, fans' loyalty, domestic league marketing and management effectiveness, as well as stadium infrastructure. Also, the variation in *Trend* allows for additional heterogeneity among countries. For such panel-dataset with $T$~50-60, it is reasonable to test for linear, quadratic and cubic linear trend.

In Model 5 all explanatory variables are assumed to have a common attendance effect for all European domestic football leagues. This assumption is reasonable since Europe is a quite homogenous continent in several aspects of social and economic life. Following the cross-country study of Nalbantis and Pawlowski (2018), football fans across leagues are expected to exhibit similar football preferences. It may be argued that the assumption of common effects across different national leagues may generate some bias due to model misspecification; however, we believe that it will be more beneficial in terms of efficiency to pool information from multiple leagues (Baltagi, Griffin, & Xiong, 2000). This claim is also supported by Attanasio, Picci, and Scorpu (2000). An additional restriction on Model 5 is that the effect of the explanatory variables remains constant over time. Such an assumption is desirable since our focus is on the long-run impact or constant effect of variables on attendance for a holistic view of competitive balance. Lags up to second order is a standard choice in order to conserve degrees of freedom for models with annual data involving a large number of parameters (Catao & Terrones, 2001). For the dataset of the current study, an initial *ADL* scheme up to the third order is tested.

The properties of *ADL* relations can better be revealed through reparameterization of the original equation in both levels and first differences (Hendry & Nielsen, 2007; Johnston & DiNardo, 1997). By switching to the reparametrized *ADL* scheme, a substantial reduction is enabled in the collinearity of the regressors, which leads to smaller standard errors of the new parameters. Coefficient estimates may be affected by the correlation between the covariate of the *Real per capita gross national disposable income (RGNI)* with the covariates of *Population* (*POP)* and the *Unemployment* (*Un)* (Borland & MacDonald, 2003). The estimated standard error of the regression, the log-likelihood values, the Durbin-Watson statistic, and the information criteria do not change (Johnston & DiNardo, 1997). The full specification of the reparametrized model is given by:

$$\begin{aligned}\Delta \ln ATT_{it} = {} & C_i + B_1(1)\ln CB_{i,t-1} + \delta_{1,0}\Delta \ln CB_{it} + \delta_{1,1}\Delta \ln CB_{i,t-1} \\ & + B_2(1)\ln POP_{i,t-1} + \delta_{2,0}\Delta \ln POP_{it} + \delta_{2,1}\Delta \ln POP_{i,t-1} \\ & + B_3(1)\ln RGNI_{i,t-1} + \delta_{3,0}\Delta \ln RGNI_{it} + \delta_{3,1}\Delta \ln RGNI_{i,t-1} \\ & + B_4(1)\ln UN_{i,t-1} + \delta_{4,0}\Delta \ln UN_{it} + \delta_{4,1}\Delta \ln UN_{i,t-1} \\ & - A(1)\ln ATT_{i,t-1} - \theta_1 \Delta \ln ATT_{i,t-1} \\ & + B_5 d97_t + \sum_{g=1}^{m} B_{(5+g)i}\, Trend^{g} + \varepsilon_{it}\end{aligned} \quad \textbf{(6)}$$

The estimation of the model using the pooled data via *OLS* tends to generate serious complications since errors may be serially correlated within cross-sectional units or countries (Hicks, 1994) as many national features (i.e. population) are not independent across years. Additionally, errors tend to be contemporaneously correlated across countries since structural factors such as the impact from TV broadcasting, the advent of advertising and sponsoring, the high-tech stadium infrastructure, and the progress in technology manufacturing football material are omitted from the equation. Lastly, errors may be heteroskedastic given the substantial difference both in size and population among the examined European countries. A common technique to improve the model is to allow for a contemporaneous correlation between error terms across equations using the *Seemingly Unrelated Regressions* (*SUR*) method, which is an *Estimated Generalised Least Squares* approach (*EGLS*) (Greene, 2008). The joint estimation of equations using the *SUR* technique accounts also for the different variances of the error terms in the equations.

The *SUR* technique is developed by Zellner (1962) and according to Kmenta and Gilbert (1968), if errors are normally distributed, iterating *SUR* yields the maximum likelihood estimates. Hill, Griffiths, and Lim (2008) propose this technique for the estimation of “long and narrow” panels while Beck and Katz (1995, 1996) propose it only if the of observed time points (here seasons) $T$ is quite large relative to $n$. They claim that only in that case, the contemporaneous variance-covariance matrix is well estimated, and the *SUR* technique improves the efficiency of estimation.

The Lagrange Multiplier (*LM*) test, suggested by Breusch and Pagan (1980), is used to identify equality of variances and zero contemporaneous correlation between errors across equations provided that the explanatory variables differ among countries. Lastly, for testing the first order autocorrelation is used the Durbin’s h-statistic for the Durbin-Watson test (Durbin & Watson, 1950, 1951) as introduced by Durbin (1970) for regression models including lagged dependent variables as explanatory variables (Asteriou & Hall, 2015).

**Empirical Results**

Using simple *OLS*, two lags of ln*ATT* are found to be significant (at $\alpha$=1%) and the first order-autocorrelation issue is solved based on the results of the Durbin’s h-statistic. Based on the *LM* test results (Table 3), the estimation of our model can be improved by employing the *SUR* technique. Our model follows the White cross-section covariance method for *SUR* models suggested by Wooldridge (2002) for the robustness of the cross-equation (contemporaneous) correlation and the heteroskedasticity (White, 1980, 1984). Using the *EGLS-SUR* method, three lags of attendance are found to be highly significant; and thus, the initial reparametrized *ADL* model in (6) is of third order. As a robustness check, we also run a fixed effects model with roughly similar results. The results, when $ADC_K$ is included in the model, are presented in Table 4. The trend results for this model are presented at the Appendix (Table 7) along with the results for all indices (Tables 8, 9, and 10) and a graphical representation of the trend per country for the attendance and selected choice of indices (Figures 2 to 7).

Depending upon the index included in the formulation, 25% to 30% of the observed variability of log-attendance is explained by the fitted model. The adjusted $R^2$ is relatively small because of two important factors: a) data of important determinants of the demand in attendance like ticket price, televised games, weather conditions, information for particular leagues or seasons were not available for this study and hence they were not included in the model, and b) the dependent variable is expressed in differences between subsequent time points. For our dataset, the correlation coefficients between the economic variables range from 0.21 to 0.39 which induces that they were no complications in the estimation of the model due to collinearity issues.

TABLE 3. *LM* Test Statistic for *SUR* Testing

| *Index in the Model* | *LM* | *Index in the Model* | *LM* |
|---|---|---|---|
| $\ln NAMSI$ | 42.859** | $\ln DN_1$ | 45.935** |
| $\ln HHI^*$ | 43.106** | $\ln DN^I$ | 42.191** |
| $\ln AGINI$ | 38.909* | $\ln ADN_K$ | 39.232* |
| $\ln NCR_1$ | 38.879* | $\ln SDN_K^I$ | 40.452* |
| $\ln NCR^I$ | 44.384** | $\ln DC_1$ | 53.774*** |
| $\ln ACR_K$ | 42.948** | $\ln DC^I$ | 46.574** |
| $\ln SRC_K^I$ | 43.916** | $\ln ADC_K$ | 39.864* |
| $\ln\tau$ | 44.053** | $\ln SDC_K^I$ | 47.361** |
| $\ln DN_t$ | 42.080** | | |

*Significant at α=10%; **significant at α=5%; ***significant at α=1%.

Concerning the basic assumptions of the model, the CIPS test results in a p-value < 1% rejecting the null hypothesis concerning the existence of unit root. The assumption concerning the normality of the residuals is not rejected based on the results of the Jarque-Bera statistic (overall p-value=0.30; all country specific p-values > 0.05; see Table 4 for details). After solving the reparametrized *ADL* Model 6 and setting all first differences to zero, the long-run elasticity effect of the explanatory variables as well as the effect for the 1997/98 indicator (dummy) are presented in Table 5. Although the sign for most of the competitive balance indices is negative, there are indices with a positive effect. The parameters of economic variables are highly significant (as expected) for α=5% level of significance. Given that residuals are stationary, there is a strong evidence of a cointegrating relation between attendance and all economic variables (Johnston & DiNardo, 1997).

The positive effect of the 1997/98 indicator provides us a reasonable interpretation of the results which is in agreement with the corresponding economic theory. Lastly, the Ramsey *RESET* test which implemented to check for the specification error has been used for the omitted variables, incorrect functional form, and correlation between explanatory variables and residuals (Ramsey & Schmidt, 1976; Ramsey, 1969). Based on the results, the *RESET* test statistic has a *p*-value higher than 0.10 for all versions of the model, which implies a well-specified model.

## 6. Analysis and Interpretation of Empirical Results

In our *ADL* model of third lag order, the overall attendance of the previous season (in log scale) and the difference of the log-attendance for year $t$ and $t$-3 have a significant effect on the difference of log-attendance between two subsequent years. With respect to population, the lagged value and the second order lagged difference are both found to be significant. As it was expected from the economic theory (Borland & MacDonald, 2003), the impact of population on attendance is found to be positive and statistically significant (p-values < 0.01); see Table 5 for the estimated effects. Moreover, the long-run elasticity is close to nine regardless of the index employed in the model); see at the second column of Table 5. For illustration, a 1% increase in national population increases football attendance almost by 9%. This result is roughly consistent with the findings from Schmidt and Berri (2001) and Scully (1989). In a similar study with domestic baseball leagues as cross units, the coefficient of population was found to be positive but not significant (Lee, 2004). Moreover, in other studies this effect was either reported as ambiguous (Coffin, 1996) or found as non-significant (Humphreys, 2002).

TABLE 4: *EGLS-SUR* Results for Attendance Model, Europe 1959/60-2018/19
Dependent Variable is $\Delta \ln ATT$

| | $\ln ADC_K$ | $\ln POP$ | $\ln RGNI$ | $Un$ | $\ln ATT$ |
|---|---|---|---|---|---|
| 1st lag: | $-0.167^{***}$ (0.035) | $3.375^{***}$ (0.744) | $0.182^{***}$ (0.059) | $-0.008^{***}$ (0.003) | $-0.403^{***}$ (0.035) |
| Δ: | $-0.119^{***}$ (0.024) | | $0.357^{***}$ (0.134) | | |
| 1st lag Δ: | | | $-0.282^{**}$ (0.128) | | |
| 2nd lag Δ: | | $-12.102^{***}$ (2.072) | | | $0.158^{***}$ (0.046) |
| | | | *d97* | *D-h*† | $R^2adj$ |
| | | | $0.033^{**}$ (0.013) | -0.921 | 0.250 |

| | Constant | Constant & Trend |
|---|---|---|
| CIPS$^a$: | $-5.846^{***}$ | $-5.959^{***}$ |
| Crit. Value α=1% | < -2 .37 | < -3.10 |

| Countries Eq.: | *BEL* | *ENG* | *FRA* | *GER* | *GRE* | *ITA* | *NOR* | *SWE* |
|---|---|---|---|---|---|---|---|---|
| *JB* (p-value) ‡: | 0.83 | 0.84 | 0.85 | 0.94 | 0.99 | 0.08 | 0.54 | 0.15 |

Total pool (unbalanced observations) after adjustment: 442

Numbers in parentheses are standard errors; Δ is the first difference.

*Significant at α=10%; **significant at α=5%; ***significant at α=1%.

†Durbin's h-statistic (value |D-h|<1.96 do not reject the null hypothesis about no autocorrelation).

‡Jarque-Bera normality test.

The long-run impact of income on attendance is considerably lower than that the corresponding one of the population and was found to be close to 0.5. To clarify, 1% increase in real per capita *GNI* brings about 0.5% increase in attendance. The magnitude of this effect is in agreement with the small *GDP* effect found by Lee (2004). Buying power has little effect on fans' decision to attend a football game. Consequently, attendance is income inelastic and not a luxury good. However, the positive coefficient suggests that attendance is a normal good, which is generally in compliance with the findings of Schmidt and Berri (2001) and Scully (1989).

Although unemployment rate is highly significant and has a negative effect on attendance, its effect is relatively low. More specifically, a constant elasticity equilibrium close to -0.15% is estimated for the current unemployment rate of ~7.5% in European Union. The sign of this effect accords with the review offered by Villar and Guerrero (2009) for a more frequent negative effect and the results of Avgerinou and Giakoumatos (2009) in their study on Greek professional football. On the contrary, Sandercock and Turner (1981) imply a positive effect justified by social factors in line with the findings of Burdekin and Idson (1991) and Falter and Perignon (2000).

The post 1997/98 season effect (captured by the binary dummy variable *d*97) is found to be significant with a positive effect on attendance. This suggests a combined effect of approximately 10% increase in attendance due to the two recent structural changes to European football; that is, the Bosman case and the Champions League reformation.

TABLE 5. Long-run Elasticity Effect of Indices and Economic Variables on Attendance; & Dummy Variable Effect

| Index in the Model | | ln*POP* | ln*RGNI* | *UN* | *d*97 |
|---|---|---|---|---|---|
| $\ln NAMSI$ | -0.093 | $9.350^{***}$ | $0.518^{***}$ | $-0.112^{**}$ | $0.115^{***}$ |
| $\ln HHI^{*}$ | -0.043 | $8.641^{***}$ | $0.517^{***}$ | $-0.112^{**}$ | $0.115^{***}$ |
| $\ln AGINI$ | 0.035 | $8.792^{***}$ | $0.518^{***}$ | $-0.102^{**}$ | $0.110^{***}$ |
| $\ln NCR_1$ | $-0.253^{***}$ | $7.463^{***}$ | $0.408^{***}$ | $-0.177^{***}$ | $0.088^{***}$ |
| $\ln NCR^I$ | $0.156^{***}$ | $8.764^{***}$ | $0.461^{***}$ | $-0.105^{**}$ | $0.094^{***}$ |
| $\ln ACR_K$ | $-0.274^{***}$ | $7.892^{***}$ | $0.512^{***}$ | $-0.123^{**}$ | $0.105^{***}$ |
| $\ln SCR_K^I$ | $-0.246^{***}$ | $8.115^{***}$ | $0.521^{***}$ | $-0.121^{**}$ | $0.109^{***}$ |
| $\ln \tau$ | -0.020 | $8.506^{***}$ | $0.502^{***}$ | $-0.103^{**}$ | $0.118^{***}$ |
| $\ln DN_t$ | $0.137^{**}$ | $8.471^{***}$ | $0.482^{***}$ | $-0.116^{**}$ | $0.106^{***}$ |
| $\ln DN_1$ | $-0.044^{***}$ | $9.220^{***}$ | $0.556^{***}$ | $-0.086^{*}$ | $0.080^{***}$ |
| $\ln DN^I$ | $0.174^{***}$ | $9.769^{***}$ | $0.426^{***}$ | $-0.147^{***}$ | $0.095^{***}$ |
| $\ln ADN_K$ | $-0.301^{***}$ | $8.789^{***}$ | $0.463^{***}$ | $-0.138^{**}$ | $0.087^{**}$ |
| $\ln SDN_K^I$ | $-0.314^{***}$ | $8.782^{***}$ | $0.558^{***}$ | $-0.102^{*}$ | $0.101^{***}$ |
| $\ln DC_1$ | $-0.303^{***}$ | $8.905^{***}$ | $0.545^{***}$ | $-0.114^{**}$ | $0.069^{**}$ |
| $\ln DC^I$ | $0.175^{***}$ | $8.842^{***}$ | $0.476^{***}$ | $-0.116^{**}$ | $0.101^{***}$ |
| $\ln ADC_K$ | $-0.414^{***}$ | $8.373^{***}$ | $0.452^{***}$ | $-0.153^{***}$ | $0.092^{**}$ |
| $\ln SDC_K^I$ | $-0.395^{***}$ | $8.526^{***}$ | $0.535^{***}$ | $-0.119^{**}$ | $0.089^{***}$ |

$^{*}$Significant at α=10%; $^{**}$significant at α=5%; $^{***}$significant at α=1%.

Lastly, with the exception of the Greek league, a cubic trend was found to be significant with an initial downward pattern until the late 70's and the mid of 80's depending upon the chosen country. Based on the results presented in Table 7 at the Appendix, trend follows a mild upward pattern until the mid or late of the 2000/10 decade and a slight downward pattern onwards. The trend variable may capture factors that affect demand for attendance that change systematically over time, such as changes in consumer preferences as far as spending their leisure time is concerned and the competition from related sports and entertainment product industry goods. An interpretation of the findings may be derived if we consider that early in 60's football in Europe was a highly respectable social phenomenon. Afterwards, however, modern forms of social events enter the entertainment industry while football remains stagnant and struggles with hooliganism. During the last three decades, the adoption of management and marketing practices both by clubs and by federations, the construction of high-tech stadiums, and the great exposure by the media have given a new noticeable boost to football. The slight downward trend detected for the last decade may be attributed as a negative effect of the prestigious European competitions to the domestic championships. The growing interest of fans for both the Champions League and the Europa League, which is verified by the steep increase of their revenues, it may be at the expense of the domestic leagues.

The *UOH* is supported only when specially designed indices for European football leagues are included in the model. In particular, the indices designed for the top *K* teams and the three-level indices are found to have a highly significant long-run elasticity with the expected sign while the magnitude of the effect considerably varies. On the other hand, the indices designed for the *I* relegated teams are also found to have a significant long-run elasticity, however, with an effect surprisingly opposite to the *UOH*; that is, the more competition on the field, the less attendance on the stadium. Most of the conventional indices, as is illustrated in Fig. 1, are not found to have a significant long-run elasticity on attendance. As an exception, only $DN_t$ index displays a significant negative effect on attendance. It is noted that conventional indices take into consideration the performance of all teams in the league.

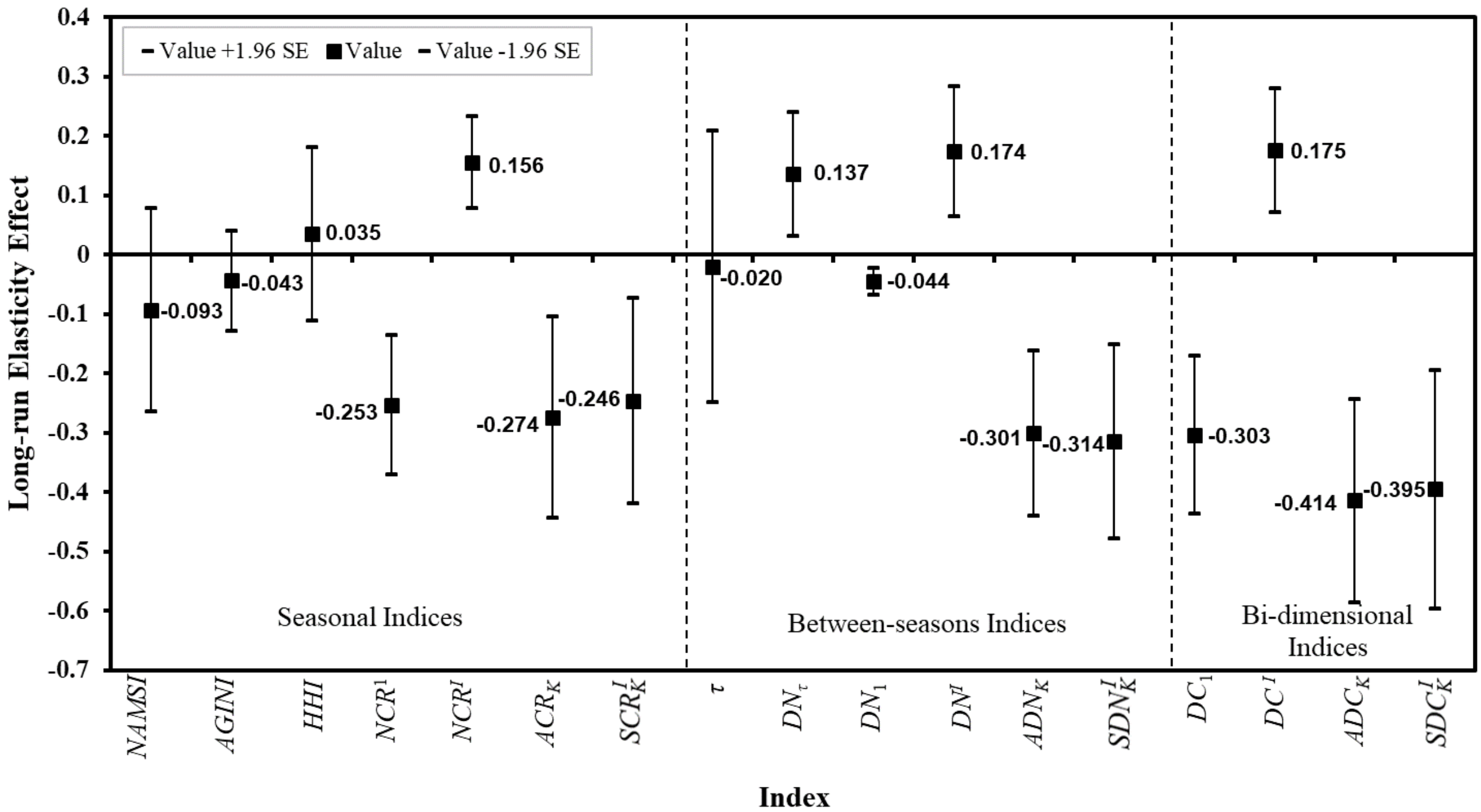

FIG 1. Long-run Elasticity and 95% Confidence Intervals
of the Effect of Competitive Balance Indices on Attendance

With concern to the seasonal indices, *NAMSI*, *HHI** and *AGINI* are not found to have a significant effect on log-attendance. Those results are consistent with the findings by Lee (2004) for a non-significant *Ratio of Standard Deviation* index, which is also a conventional index. On the other hand, Schmidt and Berri (2001) argue that the *Gini* coefficient, has a significant effect on attendance, yet only when a 3-season or a 5-season average of the index is employed in their model. One possible interpretation of these results is that, by using a conventional method of measuring seasonal competitive balance, the information gathered fails to encapsulate the fans' interest for the league.

On the other hand, the specially designed seasonal indices are found to have a highly significant effect and positively affecting attendance except for the $NCR^I$ (relegation) index. The relative weakness of the relegated teams during a particular season adversely affect the fans' behaviour. This is interpreted as more supporters at the stadium when there's less competition for relegation. The negative effect of $NCR^I$ may explains the fact that $SCR_K^I$ is found to have a lower effect than $ACR_K$, although the latter captures only two of the three levels in European football. In effect, $ACR_K$ has the greatest seasonal long-term effect which equals -0.274%. For illustration, the magnitude of that effect for the worst (2018/19) and the best seasons (1985/86) in Greece, in terms of the competitive balance values, is interpreted as a 13.8% increase in annual attendance. Given the qualities of $ACR_K$, it may be assumed that fans are mostly interested in the seasonal performance of the teams at the top of the ladder. For parsimonious reasons, $NCR_1$ may also be considered as a very important seasonal index since its long-run elasticity is very close to that of more sophisticated indices, such as $ACR_K$ and $SCR_K^I$.

Concerning the between-seasons dimension, only $\tau$ index is not found to have a significant effect on attendance. From $DN_t$ and $DN^I$ it may be inferred that the overall ranking mobility as well as the ranking mobility of the relegated teams across seasons adversely affect attendance. As for the remaining between-seasons indices, the effect of $DN_1$ is relatively small, since its long-run elasticity equals -0.044%. The magnitude of the champion's mobility effect across seasons is in sharp contrast with the corresponding effect of the champion's performance during a season. A possible explanation may be that the value of $DN_1$ equals unity in 157 out of a total of 458 cases (unity is reached, when the champion is the same for two consecutive seasons). The effect of $ADN_K$ (-0.301%), $SDN_K^I$ (-0.314%), and $DN^I$ (0.173%) is higher than the corresponding seasonal indices. Consequently, ranking mobility across seasons captures more effectively the fans' interest than seasonal performance, which accords with the findings presented from other related studies (Humphreys, 2002; Borland & MacDonald, 2003). On the contrary to the seasonal dimension, the $SDN_K^I$ displays a higher effect than the $ADN_K$ despite the opposite effect of the $DN^I$ index.

As expected from the above results, the $DC^I$ bi-dimensional index has a significant long-run opposite effect on attendance (0.175%). On the other hand, $DC_1$ and $SDC_K^I$ indices display a considerable long-run elasticity, supporting the *UOH*, with an effect of -0.303% and -395% respectively. The $ADC_K$ index has the greatest effect with a -0.414% long-run elasticity and may be considered as the best index for the study of competitive balance in European football. Attendance is negatively elastic to changes in $ADC_K$ since it increases by 0.414% for a 1% reduction in the index. From the examination of the worst (1961/62) and the best seasons (1987/88) in Greece, the effect of $ADC_K$ stands for a 15.1% increase in annual attendance or 1096 more fans to the stadium per league game. More impressively, this effect can be quantified as an increase of 6352 fans per league game for and best season (1961/62) in comparison to the worst season (2018/19) in England (Table 6).

TABLE 6. The effect of $ADC_K$ on Attendance per Country

| | *Value of* $ADC_K$ | | | | *Attendance* | |
|---|---|---|---|---|---|---|
| Country | Best Season | | Worst *Season* | | *Average*[*] | *Effect*[**] |
| Belgium | 0.530 | 1973/74 | 0.776 | 2000/01 | 9804 | 1286 |
| England | 0.373 | 1961/62 | 0.783 | 2018/19 | 29136 | 6352 |
| France | 0.311 | 1963/64 | 0.761 | 2017/18 | 14157 | 3464 |
| Germany | 0.384 | 1967/68 | 0.757 | 2013/14 | 29636 | 6043 |
| Greece | 0.517 | 1987/88 | 0.837 | 1961/62 | 6930 | 1096 |
| Italy | 0.509 | 1998/99 | 0.801 | 2017/18 | 27792 | 4192 |
| Norway | 0.277 | 1986/87 | 0.733 | 1997/98 | 6210 | 1598 |
| Sweden | 0.262 | 1967/68 | 0.718 | 1974/75 | 7635 | 2006 |

[*]The average attendance per game for the period under examination

[**]The effect of $ADC_K$ on attendance by comparing the best and the worst competitive balance seasons

The comparison among $SDC_K^I$, $ADC_K$, $DC^I$ indices allows us to further examine the role of the third level (relegation) in European football. The smaller effect of $SDC_K^I$ than the effect of $ADC_K$, might due to the inclusion of the $DC^I$ index. The absence of competition for relegation is related with higher levels of attendance. At the same time, however, when the $I$ relegated teams are less competitive, there's more competition among the rest of the teams. Actually, we can infer that the opposite to *UOH* effect of $DC^I$ is related with the effect of the other two indices. In essence, the less competition for relegation and the greater competition for the top $K$ positions are the two sides of the same coin. Yet, it is a strong argument that relegation doesn't play a significant role in European football since relegated teams usually have the smaller average attendance and stadium capacity. On the other hand, the supporters of the top $K$ teams, asking for more competition, they are bigger in numbers and greatly affect the average league attendance. The opposite effect of relegation indices might be related to the similar effect of the $DN_t$ index and the fact that the other conventional indices, which equally weigh teams at the upper and bottom half of the ladder, they are not found to have a significant effect on attendance. It is evident from the above analysis that levels of competition and ranking positions are rated in a different way by the fans.

As expected, the bi-dimensional indices have a greater effect on attendance than the corresponding seasonal and between-seasons indices; their effect is usually comparable with a set of two or even more carefully selected indices in the demand equation. The selection criteria for multiple indices on the equation are based on the correlation results and the meaningful interpretation of competitive balance. Based on the Wald test results in models with multiple indices as covariates, with the exception of the 1st level indices (measuring competitiveness in the champion level), the equality of coefficients between corresponding seasonal and between-seasons indices is not rejected at conventional levels of significance. The Wald test results for a model with four indices included simultaneously are presented at the Appendix (Tables 11 to 13).

## 7. Discussion

Following Zimbalist's (2003) suggestion, we test the uncertainty outcome hypothesis (*UOH)* in sport economics and, conditionally that *UOH* is true, we identify which measure serves best this principle and can be used for a holistic view of competitive balance in European football. Based on the principle of *UOH*, our implemented statistical model estimates the constant elasticity equilibrium among other parameters. Therefore, a number of specific sport economics theories are tested via our model-based analysis. From the model based findings, we conclude that the national population has a greater positive effect on attendance (in terms of elasticity) than the national income and the unemployment rate which are purely economic features. Another important determinant in our model was post 1997/98 season (simply included in our model by using a binary dummy variable) which indirectly captures two important events in European football: the Bosman case and the major re-structure and re-branding of the Champion's League. Finally, a cubic time trend was also found to have a significant effect on the attendance.

With regard to the competitive balance measures, our model-based empirical findings indicate that the multi-levelled indices (Manasis *et al.,* 2013; Manasis & Ntzoufras, 2014), which were specially designed to capture the peculiarities of European football leagues, were found to be significantly associated with the attendance. Therefore, the *UOH* is supported by our model implying that the more balanced is the league, the greater is the game attendance at the stadium. The results suggest the importance (in terms of fans' interest) of the first two identified levels of European football (champion level and top *K* teams) individually and in combination as reflected by the statistical significance of the combined measures. This is in line with the findings from the study of Pawlowski and Budzinski (2014) regarding the competition for the champion, the fight against relegation or other underlying sub-competitions within the championship (e.g. qualification for European tournaments). The opposite to the *UOH* effect of the indices that capture the third level of competition (relegation) is related with an even greater competition for the top *K* places. This effect might be associated with the reference-dependent preferences and loss aversion concept presented by Coates, Humphreys, and Zhou (2014). Another important finding states that the effect of between-seasons dimension (i.e. the between-season variability) is slightly higher than the corresponding seasonal one (i.e. the spread of team shares at the end of each season), which is also stated by Leeds and von Allmen (2008). However, with the exception of the 1st level (championship title), equality coefficients between the corresponding indices cannot be statistically rejected.

The two-level index which measure the competitiveness of the top *K* teams at the end of a particular season (seasonal dimension, index $ACR_K$) and the three-levelled index which measure the competitiveness across two consecutive seasons (between-seasons dimension, index $SDN_K^I$) may be considered as the most important indices expressing the *UOH* principle. However, according to our results, the best index for the study of European football (in terms of its effect on attendance) is the bi-dimensional $ADC_K$ index, which captures the first two levels in both dimensions. A considerably large effect on attendance was found for the seasonal index of champions domination. Given that competitive balance is one of the key issues that ensures the long-term success of the industry (Michie & Oughton, 2004), any conclusions derived from the analysis may be of crucial importance for key policy-makers whose aim is to sustain the viability of European football. Evidently, this effect has a considerable economic impact on total revenues both from attendance and other relates sources.

A weighting scheme of ranking places based on the specific structure of a domestic league and the stadium capacity utilization as a dependant variable both offer new area of research. Furthermore, for a more reliable estimation of the relative importance of ranking places, an empirical study could initially involve a country wise analysis using a number of country-specific variables. Subsequently, the findings concerning country-specific weighting schemes could be used for the re-estimation of competitive balance prior to a more comprehensive statistical analysis on European level, which entails a considerably large number of countries. For the estimation of the variables' coefficients we can employ random effects using hierarchical Bayesians models (Gelman & Hill, 2007) and an MCMC oriented software such as WinBUGS (Spiegelhalter, Thomas, & Best, 1998). A future research on the analysis of competitive balance (*ACB*, Fort & Maxcy, 2003) investigating break points in European football is also suggested. The *ACB* could focuses on competitive balance fluctuations due to structural changes in European football.

## Acknowledgements

This research was funded by the Research Centre of the Athens University of Economics and Business (Funding program Action 2 for the support of basic research).

We would like to thank the two referees and the editor for the constructive comments.

**Appendix**

**1. Special indices of seasonal competitive balance (Manasis *et al.,* 2013).**

a) The *Normalised Concentration Ratio for the Champion* ($NCR_1$) measures the degree champion's domination. The calculation of $NCR_1$ is based on the $NCR_K$ index (Manasis *et al.,* 2011) which measures the degree of domination degree by the top $K$ teams and is given by:

$$NCR_K = \frac{\sum_{i=1}^{K} P_i - 2K(N-1)}{2K(N-K)} = \frac{1}{2(N-K)}\left(\frac{1}{K}\sum_{i=1}^{K} P_i\right) - \frac{N-1}{N-K} \quad \textbf{(8)}$$

where $P_1$ stands for the number of points collected by the champion, $K$ stands for the number of top teams participating in European competitions the following season, and $N$ stands for the number of teams in the league. For $K$=1, the $NCR_1$ is given by:

$$NCR_1 = \frac{1}{2(N-1)} P_1 - 1 \quad \textbf{(9)}$$

The range of the index is from zero to one: zero when the champion collects 50% of the maximum attainable points, unity in the case of complete domination, in which the champion collects the maximum attainable number of points.

b) The *Adjusted Concentration Ratio* ($ACR_K$) measures two different qualities: a) the degree of domination by the top $K$ teams, and b) the degree of competition among the $K$ teams. The $ACR_K$ is derived by adjusting for the relative significance of the top $K$ positions and effectively captures both the first and the second level. The $ACR_K$ is defined as:

$$ACR_K = \frac{\sum_{i=1}^{K} NCR_i}{K} = \frac{1}{K}\left[\sum_{i=1}^{K} w_i P_i - C_K\right] \quad \textbf{(10)}$$

where $C_K$ is a constant term given by:

$$C_K = \sum_{i=1}^{K} \frac{N-1}{N-i} \quad \textbf{(11)}$$

and $w_i$ stands for the weight attached to the $i^{\text{th}}$ team given by:

$$w_i = \sum_{j=i}^{K} \frac{1}{2j(N-j)} \quad \textbf{(12)}$$

$ACR_K$ ranges from zero to one. It is zero in the absence of domination in which each of the top $K$ teams collects 50% of the maximum attainable points. In that case, the league is in a perfectly balanced state since all teams equally share points. This index is unity for both complete domination by the $K$ teams and complete imbalance among the $K$ teams. In particular, the upper bound is obtained when: a) the top $K$ teams collectively gather the maximum attainable number of points and b) within the group of $K$ top teams, any of them always wins any weaker team.

The weight $w_i$ (eq. 10, 12) which is attached to the $i$th team is derived from the partial sum of the harmonic series with first term $1/[2(N\text{-}1)]$ and last term $1/[2K(N\text{-}K)]$. Then, $w_i$ forms a sequence of the partial sums given as follows:

$$w_1 = \frac{1}{2(N-1)} + \frac{1}{4(N-2)} + \frac{1}{6(N-3)} + \ldots + \frac{1}{2K(N-K)}$$

$$w_2 = \frac{1}{4(N-2)} + \frac{1}{6(N-3)} + \ldots + \frac{1}{2K(N-K)}$$

$$w_3 = \frac{1}{6(N-3)} + \ldots + \frac{1}{2K(N-K)} \qquad \textbf{(13)}$$

...

$$w_K = \frac{1}{2K(N-K)}.$$

The first weight $w_1$ includes all the terms, the second all except the first one and so on concluding with the last weight $w_K$ which is equal to the last term of the sum given in equation (12). Each weight $w_i$ is an increasing and a decreasing function of $K$ and $N$ respectively. More importantly, from sequence (13) it can be derived that $w_i$ is a decreasing function of the ranking position, which is denoted here by index $i$. This is reasonable since the higher the ranking position (i.e. the lower $i$), the greater the interest from the fans' perspective. Furthermore, for a given $K$, the rate of this decrease in $w_i$ is an increasing function of $N$ which is also reasonable, since the champion is rated higher in a 20-team rather than in a 10-team league.

c) The *Normalised Concentration Ratio for Relegated Teams* ($NCR^I$) measure the degree of weakness of the $I$ relegated teams as compared to the remaining ones. The index is given by:

$$NCR^I = \frac{I_{PB} - \sum_{i=N-I+1}^{N} P_i}{I_{PB} - I_{UB}} = \frac{2I(N-1) - \sum_{i=N-I+1}^{N} P_i}{2I(N-1) - 2I(I-1)} = \frac{N-1}{N-I} - \frac{1}{2(N-I)}\left(\frac{1}{I}\sum_{i=N-I+1}^{N} P_i\right) \qquad \textbf{(14)}$$

where $I$ stands for the number of relegated teams, $I_{PB}$ & $I_{UB}$ stand for the maximum and minimum number of points by the I teams.

d) The *Special Concentration Ratio* ($SCR_K^I$) captures all the three levels embodied in the European multi-prized leagues. Essentially, the design of $SCR_K^I$ is based on the procedure followed for $ACR_K$. This can be simply accomplished, if $NCR^I$ is considered to be a component index of $ACR_K$. Therefore, following equations (10) and (14), $SCR_K^I$ is given by:

$$SCR_K^I = \frac{\sum_{i=1}^{K} NCR_i + NCR^I}{K+1} = \frac{1}{K+1}\left[\sum_{i=1}^{K} w_i P_i - \sum_{i=N-I+1}^{N} w_I P_i - C_K + C_I\right] \qquad \textbf{(15)}$$

where $w_I$ stands for the weight attached to the relegated teams and $C_I$ is a constant term derived from $NCR^I$ and is given by $(N\text{-}1)/(N\text{-}I)$. Similarly to the previous measures, $SCR_K^I$ ranges from zero to one. The lower bound of the index is obtained in the case where all teams share the same number of points and corresponds to a perfectly balanced league in which the indices measuring all levels of competitiveness will be constrained to their minimum values. On the other hand, the upper bound of the index is obtained when all the following conditions are simultaneously true: a) the top $K$ teams collectively gather the maximum attainable number of points, b) within the group of $K$, any team always wins against any weaker and loses from any stronger, and c) the $I$ teams collect points only from the between games. The upper bound stands for a completely unbalanced league where the indices for all levels of competitiveness will reach their maximum values.

**2) Special indices of between-seasons competitive balance (Manasis & Ntzoufras, 2014).**

a) the *Dynamic Index for the Champion* ($DN_1$) is interpreted as the degree of the champion's ranking mobility. The calculation of the index is based on the $DN_K$ index given by:

$$DN_t\text{=}1\text{-}\frac{\sum_{i=1}^{K} r_i}{K(N\text{-}K)} \tag{16}$$

where $r_i$ stands for the absolute ranking difference of the $i$th team between seasons $t$ and $t$-1. For $K$=1, the $DN_1$ is given by:

$$DN_1\text{=}1\text{-}\frac{r_1}{(N\text{-}1)} \tag{17}$$

The lower bound of $DN_1$ (zero) is obtained in the case of maximum ranking mobility, which is interpreted as the absence of dynamic domination in a league by the champion; that is, the champion comes from the last ranking place of the previous season. As far as the upper bound one is concerned, it is obtained in the case of no ranking mobility, which is interpreted as a league completely dynamically dominated by the champion; that is, the champion wins the championship for two consecutive seasons.

b) The *Adjusted Dynamic Index* ($ADN_K$) captures The index can be interpreted he degree of ranking mobility or dynamic domination by the top $K$ teams and the degree of ranking mobility or dynamic competition among the top $K$ teams. The index is given by:

$$ADN_K\text{=}\frac{\sum_{i=1}^{K} DN_i}{K}\text{=}1\text{-}\frac{2}{K}\sum_{i=1}^{K} w_i r_i \tag{18}$$

The weights $w_i$ are identical to the corresponding ones used to define the seasonal $ACR_K$. The range of $ADN_K$ accords with the conventional zero to one. The lower bound holds both for absence of dynamic domination by the top $K$ teams and perfect dynamic competition among the same teams. The lower bound is obtained in the case of maximum ranking mobility in the reverse order; that is, the top $K$ teams inversely come from the bottom of the ladder of the previous season. As the index increases, the mobility of the top $K$ teams decreases and, thus, they become more dynamically dominant. On the other hand, the upper bound stands for a dynamically completely dominated league by the top $K$ teams and absence of dynamic competition among the same teams. The upper bound is obtained when there is no ranking mobility in the top $K$ teams for two consecutive seasons.

c) the *Dynamic Index for Relegated Teams* ($DN^I$) that captures the degree of ranking mobility of the $I$ relegated teams or the degree of dynamic competition for relegation for season $t$.

$$DN^I\text{=}1\text{-}\frac{\sum_{i=N\text{-}I+1}^{N} r_i}{I(N\text{-}I)} \tag{19}$$

The lower bound zero stands for the maximum ranking mobility while the upper bound of unity stands for absence of ranking mobility.

d) the *Special Dynamic Index* ($SDN_K^I$) accounts for all three important levels in the multi-prized European football leagues. Following equations 18 and 19, the index is given by:

$$SDN_K\text{=}\frac{\sum_{i=1}^{K} DN_i\text{+}DN^I}{K\text{+}1}\text{=}1\text{-}\frac{2}{K\text{+}1}\left[\sum_{i=1}^{K} w_i r_i + \sum_{i=N\text{-}I+1}^{N} w_I r_i\right] \tag{20}$$

The zero value (lower bound) of the $SDN_K^I$ is reached for the maximum ranking mobility among the top $K$ teams as well as for the maximum ranking mobility of both the top $K$ and the bottom $I$ teams. Essentially, the top $K$ teams inversely come from the bottom $K$ positions, whereas the $I$ relegated teams come from the top $I$ positions of the previous season. The value of one (upper bound) is reached when no ranking mobility is observed in both the top $K$ and the bottom $I$ positions.

TABLE 7: Trend results for the *EGLS-SUR* model for Attendance, Europe 1959/60-2018/19, Dependent Variable is $\Delta \ln ATT$, competitive balance index in the model is the $ADC_K$

| | Belgium | England | France | Germany | Greece | Italy | Norway | Sweden |
|---|---|---|---|---|---|---|---|---|
| *Trend:* | -0.0691*** | -0.0470*** | -0.0537*** | -0.0451*** | -0.0155 | -0.0256** | -0.0984*** | -0.0827*** |
| | (0.0147) | (0.0088) | (0.0126) | (0.0121) | (0.0240) | (0.0129) | (0.0219) | (0.0147) |
| *Trend*$^2$*:* | 0.0017*** | 0.0014*** | 0.0013*** | 0.0011*** | -0.0005 | 0.0006 | 0.0027*** | 0.0021*** |
| | (0.0004) | (0.0002) | (0.0003) | (0.0003) | (0.0009) | (0.0004) | (0.0006) | (0.0004) |
| *Trend*$^3$*:* | -1.68E-05*** | -1.51E-05*** | -1.31E-05*** | -8.85E-06*** | 8.99E-06 | -7.06E-06 | -2.79E-05*** | -1.97E-05*** |
| | (3.97E-06) | (3.22E-06) | (2.99E-06) | (3.09E-06) | (1.02E-05) | (4.42E-06) | (6.33E-06) | (4.72E-06) |

TABLE 8: *EGLS-SUR* Results for Attendance Model for Seasonal Indices, Europe 1959/60-2018/19, Dependent Variable is $\Delta \ln ATT$

| | $\ln NAMSI$ | $\ln HHI^*$ | $\ln AGINI$ | $\ln NCR_1$ | $\ln NCR^I$ | $\ln ACR_K$ | $\ln SCR_K^I$ |
|---|---|---|---|---|---|---|---|
| Index | | | | | | | |
| 1$^{st}$ lag: | -0.036 | -0.017 | 0.014 | -0.104$^{***}$ | 0.062$^{***}$ | -0.110$^{***}$ | -0.097$^{***}$ |
| | (0.034) | (0.016) | (0.029) | (0.024) | (0.016) | (0.034) | (0.035) |
| Δ: | -0.046$^{**}$ | -0.023$^{**}$ | 0.037$^{**}$ | -0.087$^{***}$ | | -0.074$^{***}$ | -0.075$^{***}$ |
| | (0.022) | (0.011) | (0.017) | (0.019) | | (0.023) | (0.023) |
| 1$^{st}$ lag Δ: | 0.037$^{*}$ | 0.018$^{*}$ | | | | 0.036$^{*}$ | 0.039$^{*}$ |
| | (0.019) | (0.009) | | | | (0.021) | (0.020) |
| 2$^{nd}$ lag Δ: | | | | | | | |
| $\ln POP$ | | | | | | | |
| 1$^{st}$ lag: | 3.364$^{***}$ | 3.371$^{***}$ | 3.458$^{***}$ | 3.071$^{***}$ | 3.501$^{***}$ | 3.177$^{***}$ | 3.224$^{***}$ |
| | (0.761) | (0.761) | (0.758) | (0.725) | (0.763) | (0.726) | (0.731) |
| Δ: | | | | | | | |
| 1$^{st}$ lag Δ: | | | | | | | |
| 2$^{nd}$ lag Δ: | -10.989$^{***}$ | -10.997$^{***}$ | -10.882$^{***}$ | -10.974$^{***}$ | -10.631$^{***}$ | -11.177$^{***}$ | -11.130$^{***}$ |
| | (2.190) | (2.192) | (2.226) | (2.007) | (2.244) | (2.044) | (2.075) |
| $\ln RGNI$ | | | | | | | |
| 1$^{st}$ lag: | 0.201$^{***}$ | 0.201$^{***}$ | 0.204$^{***}$ | 0.167$^{***}$ | 0.188$^{***}$ | 0.206$^{***}$ | 0.207$^{***}$ |
| | (0.061) | (0.060) | (0.058) | (0.061) | (0.057) | (0.061) | (0.061) |
| Δ: | 0.367$^{***}$ | 0.368$^{***}$ | 0.364$^{***}$ | 0.287$^{**}$ | 0.337$^{***}$ | 0.368$^{***}$ | 0.373$^{***}$ |
| | (0.116) | (0.115) | (0.116) | (0.130) | (0.121) | (0.128) | (0.123) |
| 1$^{st}$ lag Δ: | -0.304$^{**}$ | -0.308$^{**}$ | -0.304$^{**}$ | | -0.286$^{**}$ | -0.301$^{**}$ | -0.305$^{**}$ |
| | (0.122) | (0.122) | (0.124) | | (0.130) | (0.125) | (0.122) |
| 2$^{nd}$ lag Δ: | -0.324$^{**}$ | -0.322$^{**}$ | -0.347$^{***}$ | -0.278$^{**}$ | -0.338$^{**}$ | -0.293$^{**}$ | -0.301$^{**}$ |
| | (0.134) | (0.135) | (0.130) | (0.134) | (0.138) | (0.139) | (0.138) |
| $UN$ | | | | | | | |
| 1$^{st}$ lag: | -0.006$^{**}$ | -0.006$^{**}$ | -0.005$^{**}$ | -0.010$^{***}$ | -0.005$^{**}$ | -0.006$^{**}$ | -0.006$^{**}$ |
| | (0.002) | (0.003) | (0.003) | (0.002) | (0.002) | (0.003) | (0.003) |
| Δ: | | | | | | | |
| 1$^{st}$ lag Δ: | -0.013$^{***}$ | -0.014$^{***}$ | -0.015$^{***}$ | | -0.014$^{***}$ | -0.012$^{**}$ | -0.012$^{**}$ |
| | (0.005) | (0.005) | (0.005) | | (0.005) | (0.005) | (0.005) |
| 2$^{nd}$ lag Δ: | | | | | | | |
| $\ln ATT$ | | | | | | | |
| 1$^{st}$ lag: | -0.389$^{***}$ | -0.390$^{***}$ | -0.393$^{***}$ | -0.411$^{***}$ | -0.402$^{***}$ | -0.402$^{***}$ | -0.397$^{***}$ |
| | (0.036) | (0.036) | (0.036) | (0.037) | (0.036) | (0.036) | (0.036) |
| 1$^{st}$ lag Δ: | | | | | | | |
| 2$^{nd}$ lag Δ: | 0.137$^{***}$ | 0.138$^{***}$ | 0.145$^{***}$ | 0.155$^{***}$ | 0.141$^{***}$ | 0.143$^{***}$ | 0.138$^{***}$ |
| | (0.045) | (0.049) | (0.048) | (0.048) | (0.046) | (0.049) | (0.049) |
| $d97$ | 0.045$^{***}$ | 0.045$^{***}$ | 0.043$^{***}$ | 0.036$^{***}$ | 0.048$^{***}$ | 0.042$^{***}$ | 0.043$^{***}$ |
| | (0.012) | (0.012) | (0.012) | (0.013) | (0.011) | (0.012) | (0.012) |
| $R^2 adj$ | 0.259 | 0.260 | 0.261 | 0.257 | 0.274 | 0.251 | 0.257 |

$^{*}$Significant at α=10%; $^{**}$significant at α=5%; $^{***}$significant at α=1%.

TABLE 9: *EGLS-SUR* Results for Attendance Model for Between-seasons Indices, Europe 1959/60-2018/19, Dependent Variable is $\Delta \ln ATT$

| | $\ln\tau$ | $\ln DN_t$ | $\ln DN_1$ | $\ln DN^I$ | $\ln ADN_K$ | $\ln SDN_K^I$ |
|---|---|---|---|---|---|---|
| Index | | | | | | |
| 1st lag: | -0.008 | 0.054$^{**}$ | -0.019$^{***}$ | 0.075$^{***}$ | -0.119$^{***}$ | -0.123$^{***}$ |
| | (0.045) | (0.021) | (0.005) | (0.024) | (0.028) | (0.032) |
| Δ: | -0.062$^{*}$ | | -0.068$^{**}$ | | -0.070$^{***}$ | -0.082$^{***}$ |
| | (0.022) | | (0.003) | | (0.017) | (0.020) |
| 1st lag Δ: | | | 0.008$^{***}$ | -0.042$^{**}$ | | |
| | | | (0.002) | (0.020) | | |
| 2nd lag Δ: | | | 0.004$^{***}$ | -0.032$^{*}$ | | |
| | | | (0.001) | (0.017) | | |
| $\ln POP$ | | | | | | |
| 1st lag: | 3.348$^{***}$ | 3.387$^{***}$ | 4.137$^{***}$ | 4.222$^{***}$ | 3.488$^{***}$ | 3.452$^{***}$ |
| | (0.734) | (0.754) | (0.814) | (0.736) | (0.755) | (0.723) |
| Δ: | | | | | | |
| 1st lag Δ: | | | | | | |
| 2nd lag Δ: | -10.599$^{***}$ | -10.657$^{***}$ | -13.534$^{***}$ | -13.811$^{***}$ | -11.783$^{***}$ | -11.467$^{***}$ |
| | (2.236) | (2.266) | (2.380) | (2.519) | (2.148) | (2.099) |
| $\ln RGNI$ | | | | | | |
| 1st lag: | 0.197$^{***}$ | 0.192$^{***}$ | 0.249$^{***}$ | 0.184$^{***}$ | 0.183$^{***}$ | 0.219$^{***}$ |
| | (0.058) | (0.057) | (0.064) | (0.058) | (0.058) | (0.061) |
| Δ: | 0.357$^{***}$ | 0.363$^{***}$ | 0.363$^{***}$ | 0.373$^{***}$ | 0.366$^{***}$ | 0.332$^{***}$ |
| | (0.116) | (0.121) | (0.126) | (0.114) | (0.132) | (0.123) |
| 1st lag Δ: | -0.305$^{**}$ | -0.316$^{**}$ | -0.344$^{**}$ | -0.281$^{**}$ | -0.304$^{**}$ | -0.316$^{**}$ |
| | (0.133) | (0.131) | (0.133) | (0.119) | (0.127) | (0.128) |
| 2nd lag Δ: | -0.358$^{***}$ | -0.378$^{***}$ | -0.319$^{**}$ | -0.358$^{***}$ | | -0.278$^{**}$ |
| | (0.132) | (0.135) | (0.143) | (0.134) | | (0.135) |
| $UN$ | | | | | | |
| 1st lag: | -0.005$^{**}$ | -0.006$^{**}$ | -0.005$^{*}$ | -0.008$^{***}$ | -0.007$^{***}$ | -0.005$^{*}$ |
| | (0.002) | (0.002) | (0.002) | (0.002) | (0.003) | (0.002) |
| Δ: | | | | | | |
| 1st lag Δ: | -0.014$^{***}$ | -0.015$^{***}$ | -0.010$^{**}$ | -0.012$^{**}$ | | -0.011$^{**}$ |
| | (0.005) | (0.005) | (0.004) | (0.004) | | (0.005) |
| 2nd lag Δ: | | | | | | |
| $\ln ATT$ | | | | | | |
| 1st lag: | -0.393$^{***}$ | -0.399$^{***}$ | -0.448$^{***}$ | -0.432$^{***}$ | -0.396$^{***}$ | -0.393$^{***}$ |
| | (0.035) | (0.035) | (0.037) | (0.036) | (0.035) | (0.036) |
| 1st lag Δ: | | | | | | |
| 2nd lag Δ: | 0.150$^{***}$ | 0.151$^{***}$ | 0.142$^{***}$ | 0.119$^{**}$ | 0.167$^{***}$ | 0.168$^{***}$ |
| | (0.047) | (0.048) | (0.046) | (0.047) | (0.046) | (0.046) |
| *d97* | 0.046$^{***}$ | 0.042$^{***}$ | 0.036$^{***}$ | 0.041$^{***}$ | 0.033$^{***}$ | 0.039$^{***}$ |
| | (0.012) | (0.012) | (0.011) | (0.011) | (0.013) | (0.012) |
| $R^2 adj$ | 0.254 | 0.255 | 0.281 | 0.293 | 0.242 | 0.251 |

$^{*}$Significant at α=10%; $^{**}$significant at α=5%; $^{***}$significant at α=1%.

TABLE 10: *EGLS-SUR* Results for Attendance Model for Bi-dimensional Indices, Europe 1959/60-2018/19, Dependent Variable is $\Delta \ln ATT$

| | $\ln DC_1$ | $\ln DC^I$ | $\ln ADC_K$ | $\ln SDC_K^I$ |
|---|---|---|---|---|
| Index | | | | |
| 1st lag: | -0.126$^{***}$ | 0.070$^{***}$ | -0.167$^{***}$ | 0.156$^{***}$ |
| | (0.028) | (0.021) | (0.035) | (0.040) |
| Δ: | -0.077$^{***}$ | | -0.119$^{***}$ | -0.133$^{***}$ |
| | (0.016) | | (0.024) | (0.027) |
| 1st lag Δ: | 0.026$^{**}$ | | | |
| | (0.012) | | | |
| 2nd lag Δ: | | | | |
| $\ln POP$ | | | | |
| 1st lag: | 3.703$^{***}$ | 3.546$^{***}$ | 3.375$^{***}$ | 3.374$^{***}$ |
| | (0.780) | (0.756) | (0.744) | (0.716) |
| Δ: | | | | |
| 1st lag Δ: | | | | |
| 2nd lag Δ: | -12.207$^{***}$ | -10.895$^{***}$ | -12.102$^{***}$ | -11.571$^{***}$ |
| | (2.089) | (2.291) | (2.072) | (2.043) |
| $\ln RGNI$ | | | | |
| 1st lag: | 0.227$^{***}$ | 0.191$^{***}$ | 0.182$^{***}$ | 0.211$^{***}$ |
| | (0.063) | (0.058) | (0.059) | (0.062) |
| Δ: | 0.359$^{***}$ | 0.383$^{***}$ | 0.357$^{***}$ | 0.328$^{***}$ |
| | (0.130) | (0.120) | (0.134) | (0.123) |
| 1st lag Δ: | -0.350$^{**}$ | -0.293$^{**}$ | -0.282$^{**}$ | -0.298$^{**}$ |
| | (0.135) | (0.123) | (0.128) | (0.126) |
| 2nd lag Δ: | -0.254$^{*}$ | -0.365$^{***}$ | | -0.244$^{*}$ |
| | (0.143) | (0.137) | | (0.139) |
| $UN$ | | | | |
| 1st lag: | -0.006$^{**}$ | -0.006$^{**}$ | -0.008$^{***}$ | -0.006$^{***}$ |
| | (0.002) | (0.002) | (0.003) | (0.003) |
| Δ: | | | | |
| 1st lag Δ: | -0.010$^{**}$ | -0.013$^{***}$ | | -0.011$^{***}$ |
| | (0.005) | (0.005) | | (0.005) |
| 2nd lag Δ: | | | | |
| $\ln ATT$ | | | | |
| 1st lag: | -0.415$^{***}$ | -0.401$^{***}$ | -0.403$^{***}$ | -0.396$^{***}$ |
| | (0.036) | (0.036) | (0.035) | (0.036) |
| 1st lag Δ: | | | | |
| 2nd lag Δ: | 0.155$^{***}$ | 0.145$^{***}$ | 0.158$^{***}$ | 0.159$^{***}$ |
| | (0.047) | (0.048) | (0.046) | (0.046) |
| $d97$ | 0.029$^{**}$ | 0.045$^{***}$ | 0.033$^{**}$ | 0.040$^{***}$ |
| | (0.012) | (0.011) | (0.013) | (0.012) |
| $R^2adj$ | 0.267 | 0.297 | 0.250 | 0.257 |

$^{*}$Significant at α=10%; $^{**}$significant at α=5%; $^{***}$significant at α=1%.

TABLE 11: *EGLS-SUR* Results for Attendance Model for Bi-dimensional Indices, Europe 1959/60-2018/19, Dependent Variable is Δln*ATT*

| | ln$NCR^I$ | ln*ACRK* | ln$DN^I$ | ln$ADN^K$ | ln*POP* | ln*RGNI* | *UN* | ln*ATT* |
|---|---|---|---|---|---|---|---|---|
| 1st lag: | 0.039$^{*}$ | -0.107$^{***}$ | 0.068$^{***}$ | -0.163$^{***}$ | 4.313$^{***}$ | 0.222$^{***}$ | -0.010$^{***}$ | -0.432$^{***}$ |
| | (0.023) | (0.034) | (0.024) | (0.044) | (0.712) | (0.062) | (0.003) | (0.037) |
| Δ: | | -0.064$^{**}$ | | -0.053$^{***}$ | | 0.423$^{***}$ | | |
| | | (0.025) | | (0.019) | | (0.109) | | |
| 1st lag Δ: | 0.030$^{**}$ | | -0.038$^{*}$ | 0.047$^{*}$ | | -0.250$^{**}$ | -0.009$^{**}$ | |
| | (0.015) | | (0.020) | (0.028) | | (0.104) | (0.004) | |
| 2nd lag Δ: | | | -0.029$^{*}$ | 0.048$^{**}$ | 15.342$^{***}$ | -0.245$^{*}$ | | 0.090$^{**}$ |
| | | | (0.016) | (0.020) | (2.191) | (0.139) | | (0.046) |
| *d97* | 0.037$^{***}$ | | | | | | | |
| | (0.010) | | | | | | | |
| $R^2adj$ | 0.336 | | | | | | | |

$^{*}$Significant at α=10%; $^{**}$significant at α=5%; $^{***}$significant at α=1%.

TABLE 12: *WALD* test Results
$H_0$: C(*ln*($NCR^I$(-1)) = C(ln($DN^I$(-1))

| Test Statistic | Value | df | Probability |
|---|---|---|---|
| t-statistic | 0.822 | 381 | 0.41 |
| F-statistic | 0.676 | (1, 381) | 0.41 |
| Chi-square | 0.676 | 1 | 0.41 |

TABLE 13: *WALD* test Results
$H_0$: C(*ln*($ACR_K$(-1)) = C(ln($ADN_K$(-1))

| Test Statistic | Value | df | Probability |
|---|---|---|---|
| t-statistic | 0.935 | 381 | 0.35 |
| F-statistic | 0.874 | (1, 381) | 0.35 |
| Chi-square | 0.874 | 1 | 0.34 |

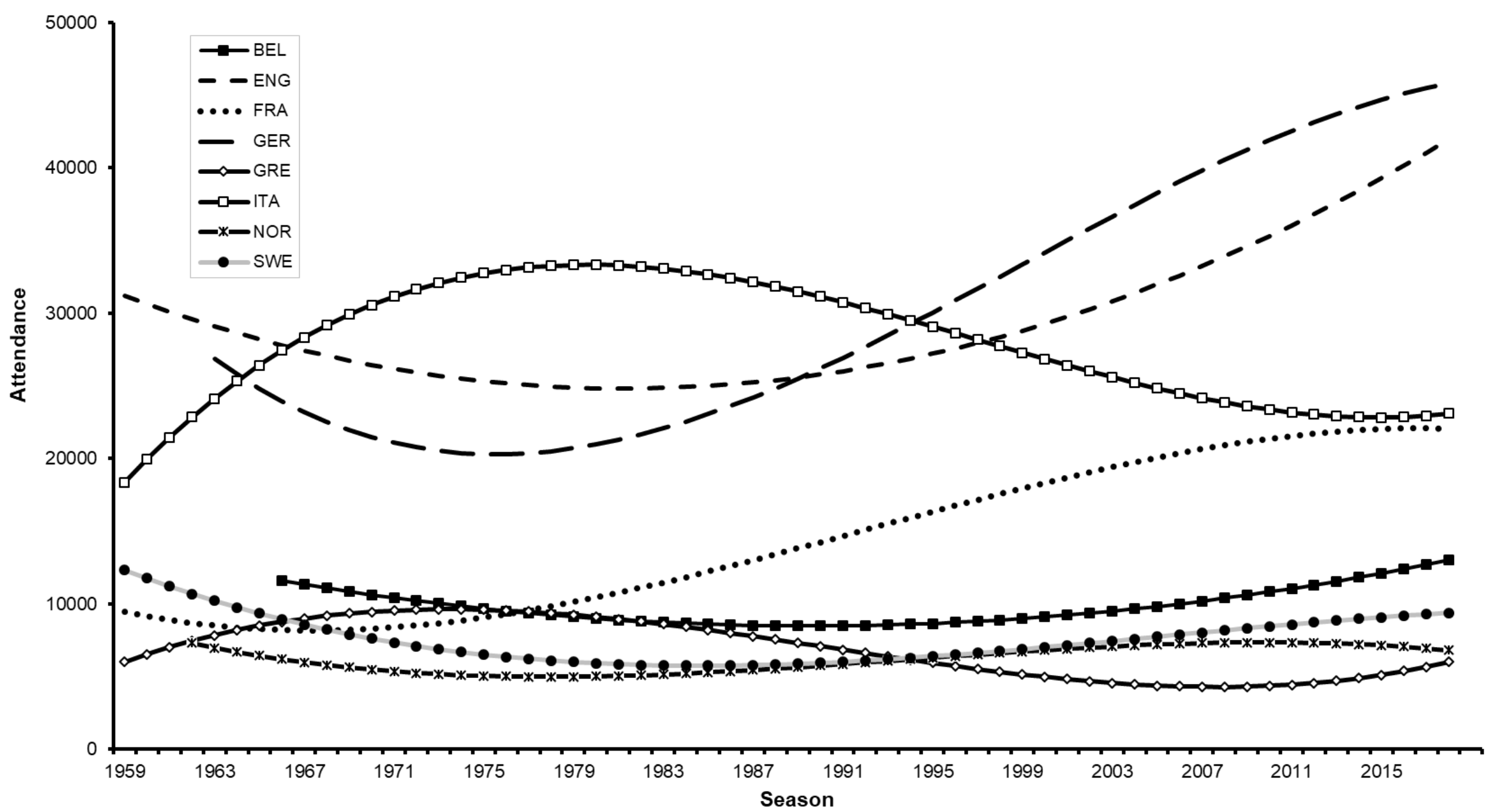

FIG 2. Trend in attendance per country

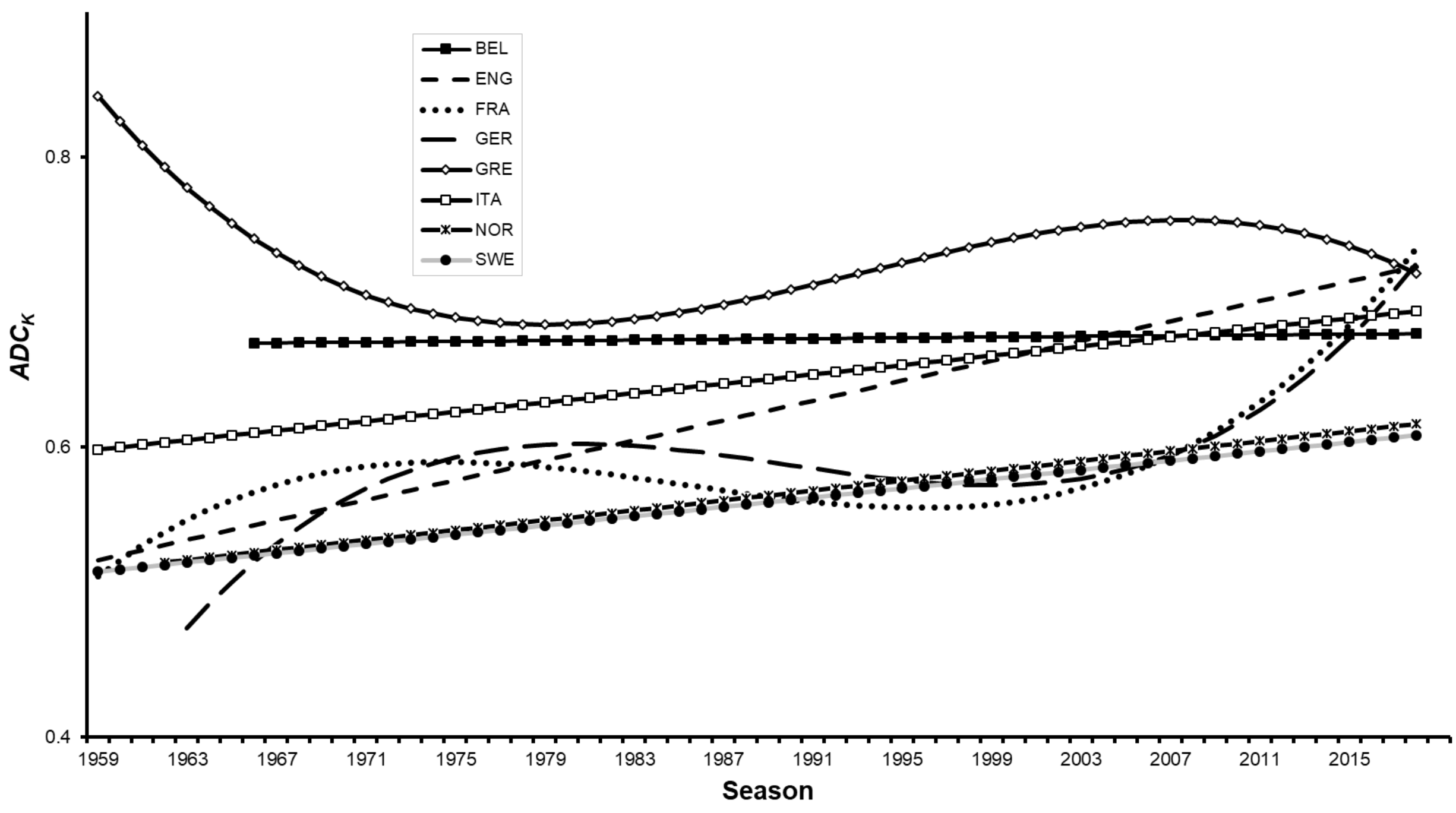

FIG 3. Trend in $ADC_K$ index per country

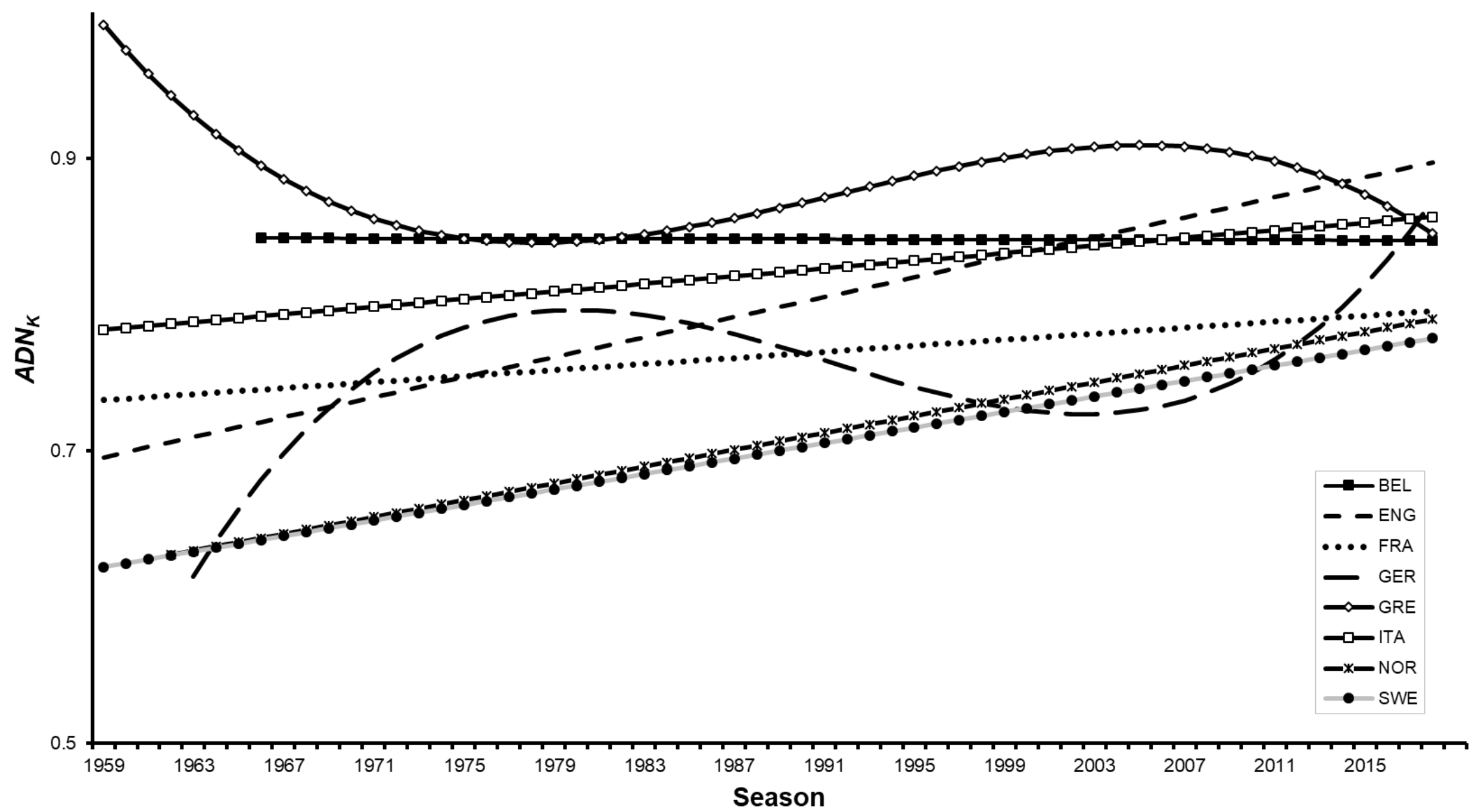

FIG 4. Trend in $ADN_K$ index per country

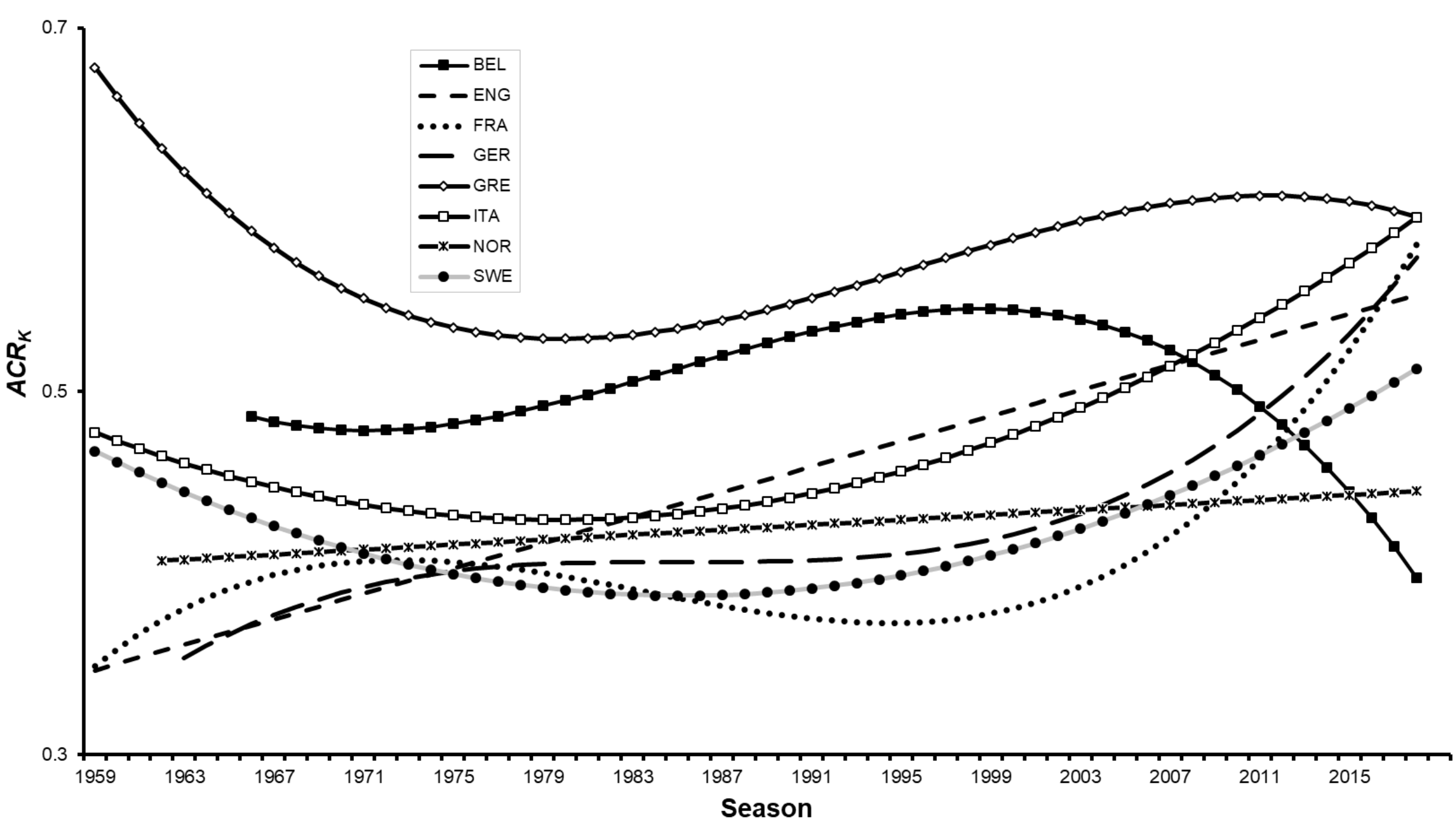

FIG 5. Trend in $ACR_K$ index per country

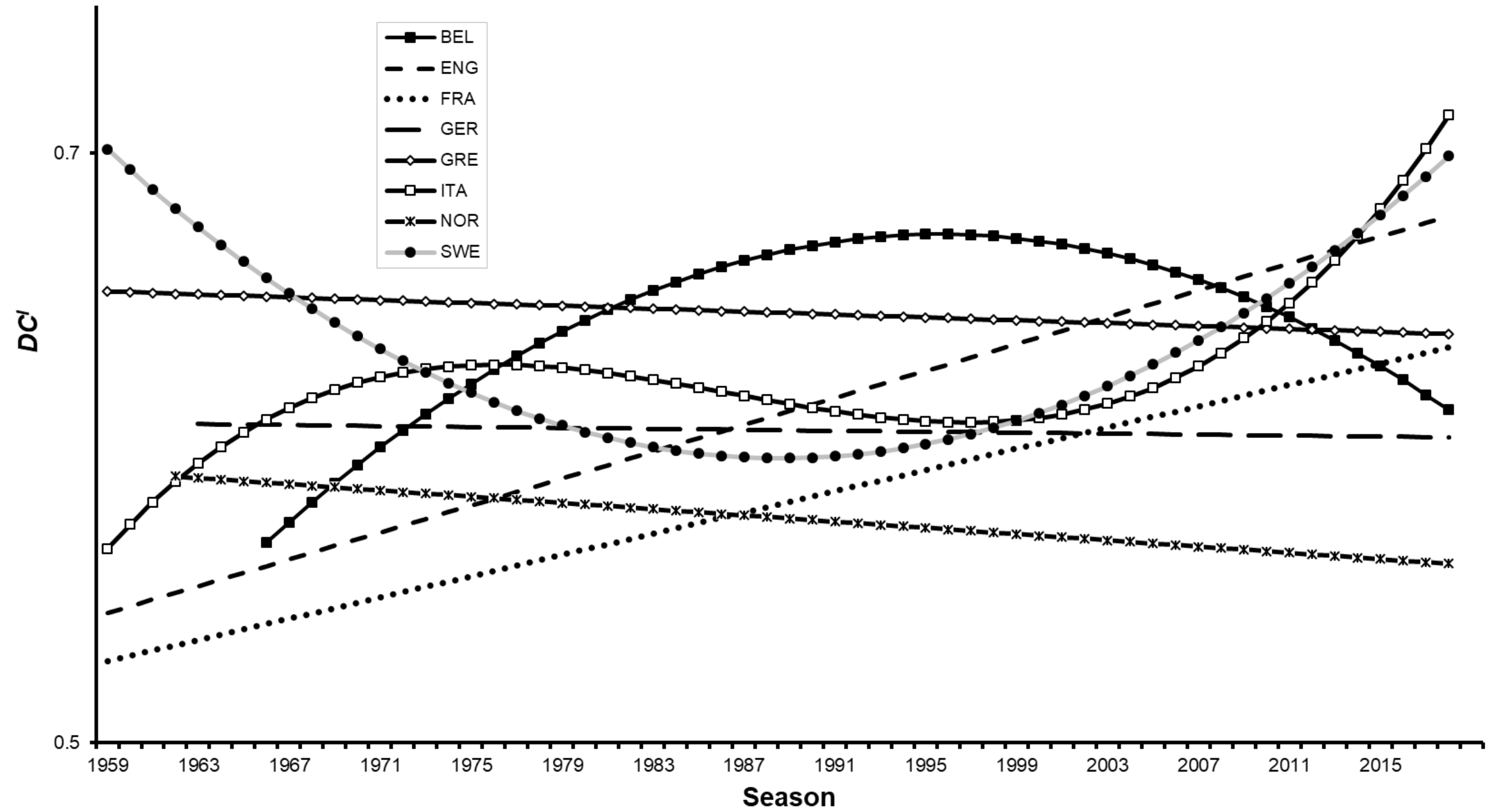

FIG 6. Trend in $DN^l$ index per country

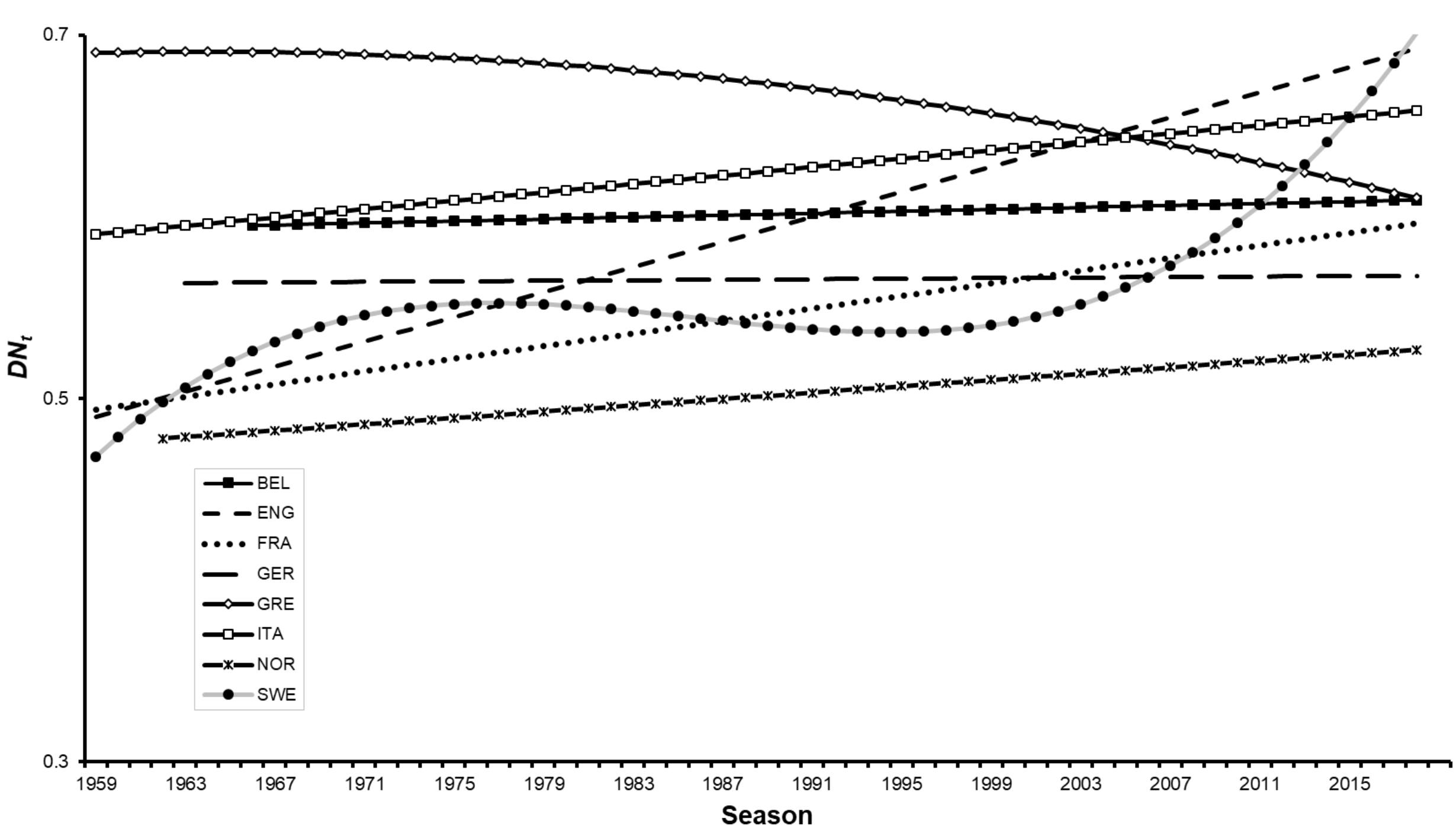

FIG 7. Trend in $DN_t$ index per country